\documentclass[12pt]{article}
\usepackage{float}
\usepackage{amsmath,amssymb}
\usepackage{mathrsfs}
\usepackage{braket}
\usepackage[dvipdfmx]{color}
\usepackage{framed,}
\usepackage[dvipdfmx,hiresbb]{graphicx}

\usepackage{bm}
\usepackage{comment}
\usepackage{tikz}
\usepackage{subcaption}

\def\nonu{\nonumber}

\newcommand{\pa}{  \partial }

\makeatletter

\renewcommand\section{\@startsection {section}{1}{\z@}%
                                   {-3.5ex \@plus -1ex \@minus -.2ex}%
                                   {2.3ex \@plus.2ex}%
                                   {\normalfont\large\bfseries}}

\renewcommand\subsection{\@startsection{subsection}{2}{\z@}%
                                     {-3.25ex\@plus -1ex \@minus -.2ex}%
                                     {1.5ex \@plus .2ex}%
                                     {\normalfont\normalsize\bfseries}}

\makeatother

\usepackage[driverfallback=dvipdfmx]{hyperref}

\begin{document}

\baselineskip=18pt  
\numberwithin{equation}{section}  
\allowdisplaybreaks  



%
%


\thispagestyle{empty}

\vspace*{-2cm}
\begin{flushright}
\end{flushright}

\begin{flushright}
\end{flushright}

\begin{center}

\vspace{1.4cm}

{\bf \Large Life-time of Metastable  Vacuum in String Theory  }
\vspace*{0.2cm}

{\bf\Large and  Trans-Planckian Censorship Conjecture }

\vspace{1.3cm}

{\bf
 Sohei Tsukahara} \footnote{E-mail:~tsukahara.sohei.256@s.kyushu-u.ac.jp} \\
\vspace*{0.5cm}

{\it Department of Physics, Kyushu University, Fukuoka 810-8581, Japan  }\\

\vspace*{0.5cm}

\vspace*{0.5cm}

\end{center}

\vspace{1cm} \centerline{\bf Abstract} \vspace*{0.5cm}

It has been known that the catalytic effect makes the life-time of a metastable state shorter. We discuss this phenomenon in a decay process of a metastable vacuum in the brane-limit of type IIB string theory. Due to the non-linear effect of DBI action, the bubble created by the decay makes an energetically favorable bound state with an impurity that plays the role of catalyst, which is quite specific to this model and different from other catalysts such as a back hole. Furthermore, we found that this low-energy effective theory around almost unstable regions reduces to a simple quantum mechanical system, and the vacuum life-time can be calculated using known results, even beyond the WKB approximation. Finally, we compare the life-time of the vacuum with the Trans-Planckian Censorship Conjecture (TCC) and find that as long as the string scale is at least one order magnitude smaller than the Planck scale, there is a nonzero window to satisfy the TCC condition.

\newpage
\setcounter{page}{1} 

\newpage
\setcounter{page}{1} 



\tableofcontents

\section{Introduction}

The trans-Planckian censorship conjecture (TCC), which was initially discussed in a pure gravity theory \cite{Original}, has attracted renewed attention in string theories in the context of the swampland program \cite{TCC1,TCC2,TCCreview}. The swampland program, initiated by \cite{SW1,SW2,SW3,SW4,SW5,SW6}, offers a set of conditions coming from empirical experiences in string theories (see \cite{SWreview} for reviews). Seemingly different conditions are interrelated and provide relations between consistency conditions constraining low-energy effective theories. In \cite{TCC1,TCC2}, the authors incorporated the TCC bound and the swampland program and studied the implications of TCC to the low energy effective theories originating from string theories.

TCC forbids enlarging a fluctuation of sub-Planckian scale to that of a classical scale during inflation. Hence, this condition imposes a strong constraint on the vacuum structure of gravity theories, especially string theories. In \cite{TCC1,TCC2}, the authors studied the life-time of de Sitter vacua without matter using Coleman-de Luccia type solutions. They showed that TCC is quite strong, and most of the de Sitter vacua in the Planck scale size of vacuum energy were excluded.

On the other hand, in nature, it is widely known that some impurities promote phase transitions, so we have room for introducing some matters as catalysts to enhance the instability.
The idea of catalysis was introduced in field theories first \cite{Steinhardt,Hosotani:1982ii,Yajnik:1986tg}, and later gravity theories\footnote{One of the remarkable example of catalysis in gravity theory can be seen in the use of singularities on the bounce solution, which was recently developed by \cite{Gregory:2013hja}.}. 
This idea was also used in phenomenological model buildings in string theory. Well-studied examples are a D3-brane catalyst \cite{Kasai:2015exa,Kasai:2015maa,Nakai:2018hhf}, which behaves as a solitonic impurity. Such D3-brane enhances the instability of metastable vacuum and induces inhomogeneous vacuum decay.

By considering the catalytic effect, a stringy vacuum naively excluded from the landscape via TCC condition can circumvent the swampland thanks to some impurities. In this paper, we engineered a metastable vacuum by wrapping D5-branes and anti D5-branes around conical singularities, which was first discussed in \cite{Aganagic:2006ex}, and considered a situation in that D3-brane impurity induces inhomogeneous vacuum decay. This model was originally introduced in \cite{Kasai:2015exa}. While they assumed to nucleate a relatively large bubble, here we adopt Dirac-Born-Infeld (DBI) action to do rigorous calculations when the bubble is small. We will study the influence of the monopole on the tunneling rate by WKB analysis and numerical calculations, including a 1-loop fluctuation factor. When the potential barrier is low and the WKB approximation breaks down, the calculations are reduced to a cubic oscillator, which allow us to compare with TCC condition.

The organization of this paper is as follows: In section 2, we first review the trans-Planckian censorship conjecture and upper bound of life-time on metastable de Sitter vacua. In section 3, we review our set-up with Dp-branes compactified on the generalized conifold and calculate the decay rate of the metastable state via the bounce solution of Euclidean DBI action. Here We show a complete 1-loop WKB result, including the determinant prefactor. In section 4, we examine the validity of the 1-loop approximation and indicate that the life-time calculation breaks down near a critical point where the potential barrier vanishes. We approximate the action in this region by a cubic oscillator, showing that the life-time is finite even when the potential is perfectly flat. Finally, we compare the minimal life-time obtained here with the TCC condition to obtain a bound on the parameter space. Section 5 is devoted to a summary and discussion. Appendices provide a detailed derivation of our action and some technical ingredients.

\section{Review of Trans-Planckian censorshop conjecture}

This paper discusses the implications of the trans-pPanckian censorship conjecture to a metastable vacuum with some non-trivial matters in type IIB string theory. Toward this goal, we first review TCC and show a general constraint on the life-time coming from the TCC, which will eventually be applied to our specific model. 

Brandenberger and Martin originally proposed TCC in \cite{Original}. As the inflation extends a quantum fluctuation to the classical one, if it lasts longer, fluctuations smaller than the Planck length can get larger than the Hubble scale $H^{-1}$. The sub-Planckian scale depends on details of quantum gravity theory, so this conjecture is related to quantum gravity. Recently, in string theory, the swampland program attracted much attention, and TCC bound was revisited in this context \cite{TCC1,TCC2}. According to the papers, the TCC bound can be translated into the following swampland condition, which has to satisfy low energy effective theories in the FRW space-time, 
\begin{equation}
{a_f \over a_i}< {M_{\rm pl}\over H_f} ~, \label{TCCcond}
\end{equation}
where $a_{i}$ and $a_f$ are the scale factor of the space-time at the initial and final state in a time interval. $H_f$ is the Hubble scale at the final state. Applying this condition to the inflationary period, one can express it as a condition for the inflation time,
\begin{eqnarray}
\tau \le H_I^{-1}  \log {M_{\rm pl} \over H_I}~,\qquad H_I=\sqrt{8\pi G V_{I}\over 3},
\end{eqnarray}
where $H_I$ is the Hubble parameter during the inflation.
The life-time $\tau$ of the metastable state is naively estimated by using WKB approximation and the bounce action $B$ of the decay process \cite{Coleman:1977py},  
\begin{eqnarray}
\tau^{WKB} = \Big(\sqrt{2\pi \over B} \Big)^{n_{\rm zero}} F(R_i ,R_f)\, e^B ~, \label{NaiveFormula}
\end{eqnarray}
where $n_{\rm zero}$ is the number of zero-modes around the bounce solution. $F(R_i , R_f)$ comes from a one-loop determinant of the path integral and is typically not so easy to calculate the exact form. However, as we will discuss, this leading approximation cannot apply to the very flat potential where the vacuum becomes almost unstable. Regarding catalytic decay, modifying the contribution from the zero modes is necessary. Specifically, we need to clarify how these modifications will affect the process.

\section{Semi-classical vacuum decay in the context of $D$-brane}

\subsection{Setup of metastable state in Type IIB string theory}

Here, we quickly review the metastable state discussed in \cite{Cachazo:2001jy,Vafa:2000wi,Aganagic:2006ex}, as an excellent example to show our idea of catalytic decay in string theory. Let us introduce D5 and anti D5-branes and partially wrap them on the two-cycles in the internal space. Here we show some results which will be used in this section, leaving the derivation of effective action and the geometry setup and so on in the Appendix A. Roughly, the internal geometry has a set of two-cycles which create the potential barrier between D5 and anti D5-branes. As is estimated by using thin-wall approximation, the metastable vacuum is long-lived as long as the following condition is satisfied,
\begin{eqnarray}
B_{O(4)}={27\pi^2 \over 2 }{T_{DW}^4\over (\Delta V)^3} \gg 1,
\label{BO4}
\end{eqnarray}
where $V$ is the difference in the energy density between a metastable vacuum and that of a true one. As we will see below, the catalytic effect triggered by a D3-brane makes the life-time much shorter.  

The decay process of this system can be understood from a four-dimensional point of view: In decaying the vacua, D5-brane and anti D5-branes annihilate, and a spherically-symmetric bubble will be generated, which the true vacuum occupies inside of the wall. We can consider this bubble as a D5-brane wrapping on the internal space, whose geometry is $S^3$, and also as a throat connecting those two branes. Although the domain wall has tension, which tends to reduce the size of bubble, this system gains energy when the inside of the bubble becomes a true vacuum. If the tunneling effect generates a large enough bubble, the bubble expands, and the metastable vacuum becomes a true vacuum. That is all description of the decay process with no catalysis.

\begin{figure}[H]
\centering
\hfill
\begin{minipage}[htbp]{0.45\linewidth}
    \begin{center}
    \includegraphics[width=0.9\linewidth]{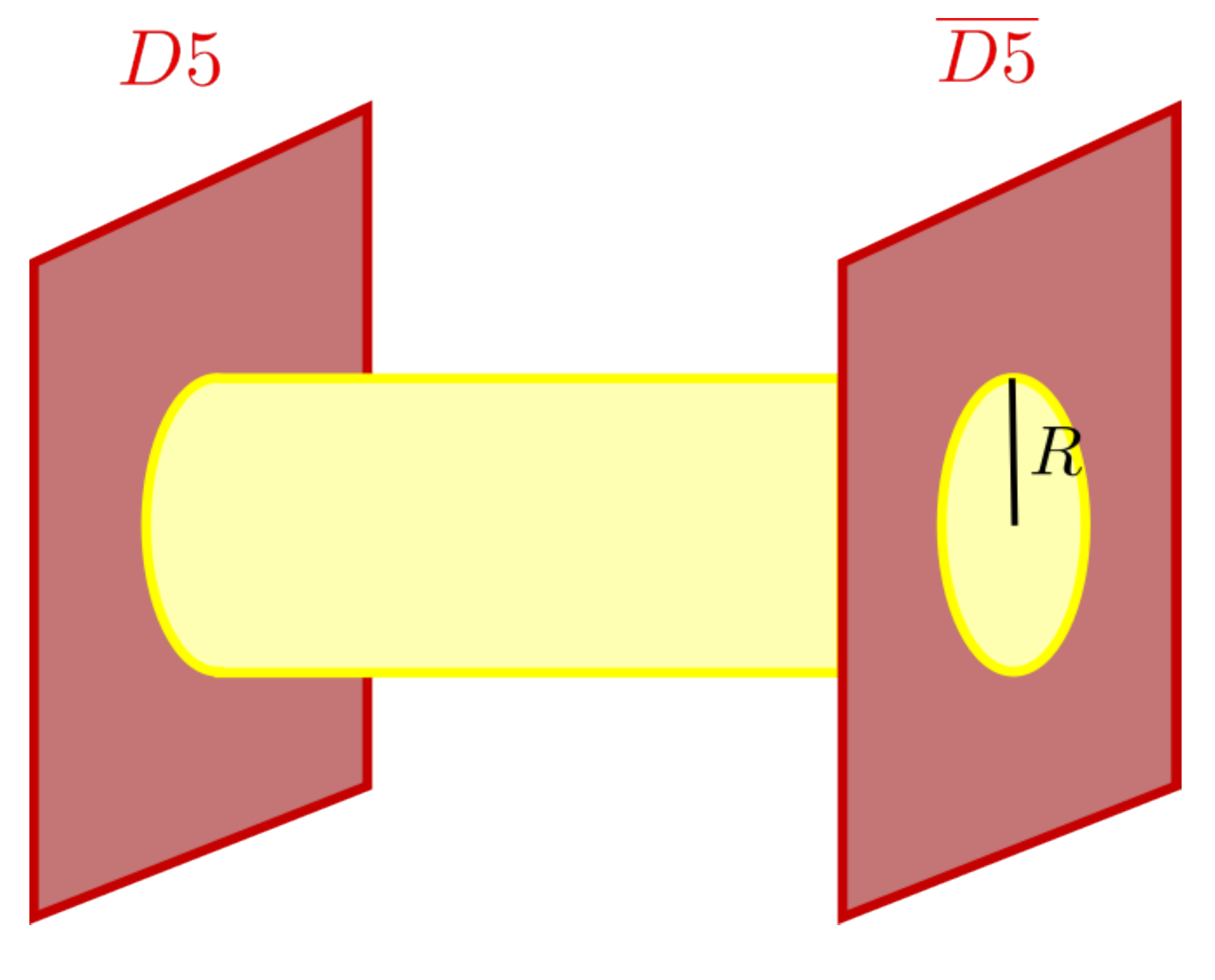}
    \vspace{-.0cm}
\end{center}
\end{minipage}
\begin{minipage}[htbp]{0.45\linewidth}
    \begin{center}
    \includegraphics[width=0.65\linewidth]{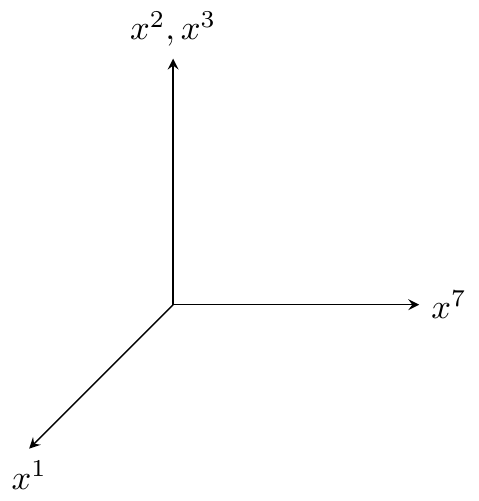}
    \vspace{-.0cm}
    \end{center}
\end{minipage}
\caption{\sl The schematic diagram for annihilation of D5-branes and anti-D5-branes.
    }
\end{figure}

\subsection{Inclusion of D3-brane and effective action for D5-D3 system}

We then introduce a catalytic D3-brane to the above system by wrapping it on $S^3$. This brane can be seen as a monopole-like object from 4d space-time \cite{StringText1,StringText2} and build a bound-state with the bubble. This is due to Dp-brane's property, which reduces the total energy via making the bound-state. The bound state energy can be lower than the simple sum \cite{StringText1,StringText2}
\begin{align}
E_{D3}+E_{D5} > E_{\rm bound}=\sqrt{E_{D3}^2+E_{D5}^2}.
\end{align}
RHS is the bound state energy and reflects a nonlinearity from DBI action. Given string ``dissolving,''  we can deal with the remnant of the D3-brane on the D5-brane as a background magnetic field.

We now show the specific expression of the Lagrangian, and the derivation is postponed to Appendix A. In this system, the metastable state decays through the nucleation of the bubble in the Minkowski space. If we write $t_{E}$ as a Euclidean time to calculate the decay rate, we can write down the Euclidean action for the bound state which determines the time dependence of the radius $R$, as
\begin{eqnarray}
L_{\rm total}= T_{DW} 4\pi   \sqrt{(R^4+b_{D3}^2)(1+\dot{R}^2)}-\Delta V  b_{D3} \Big[4\pi R\cdot {}_2F_{1} \Big(-{1\over 2}, {1\over 4}, {5\over 4},-{R^4\over b_{D3}^2} \Big) \Big]+\cdots~, \label{DBIlag}
\end{eqnarray}
where ellipses are any terms independent of $R$. The magnetic field $b_{D3}$ on the D5-brane is the trace of the D3-brane, which dissolved into the D5-brane, then its magnitude is proportional to the number of D3-branes, $b_{D3}\propto \#_{D3}$. Here, $T_{DW}$, which is the tension of the bubble that appeared in Minkowski space through the decay process between D5-branes and anti D5-branes, is given by $T_{D5}V_3$, where $V_{3}$ is the volume of the 3-cycle, that the internal space D5-branes occupy. In this Lagrangian, $r$ is an integrated value for background NS B-field in the internal space, $r=\oint_{{S}^2} B^{NS}$. We then introduce dimensionless parameters to see the dependence on the magnetic field later,
\begin{eqnarray}
R=c \widetilde{R} \ , \quad b_{D3}=c^2 \tilde{b}_{D3} \  , \quad t_E =cs\ , \quad c={T_{DW}\over \Delta V}.
\end{eqnarray}
We get the dimensionless Lagrangian with the above parameters,
\begin{gather}
\begin{gathered}
\widetilde{L}= \sqrt{(\widetilde{R}^4+\tilde{b}_{D3}^2)(1+\dot{\widetilde{R}}{}^2)}- \tilde{b}_{D3}\widetilde{R} \cdot {}_2F_{1} \Big(-{1\over 2}, {1\over 4}, {5\over 4},-{\widetilde{R}^4\over \tilde{b}_{D3}^2} \Big) +\cdots,  \\
S=\int ds A_B \widetilde{L}\ , \qquad A_B={4\pi T_{DW}^4 \over \Delta V^3}.
\end{gathered}
\label{DBIlag_nondim}
\end{gather}
Static energy is given by
\begin{equation}
\tilde{E}=\sqrt{\widetilde{R}^4+\tilde{b}_{D3}^2}-\tilde{b}_{D3}\widetilde{R}{}_{2}F_{1}\left(-\frac{1}{2},\frac{1}{4},\frac{5}{4},-\frac{\widetilde{R}^4}{\tilde{b}_{D3}^2} \right).
\end{equation}
As seen from Figure 2, as the magnetic field increases, instability of the system is enhanced, and the potential barrier vanishes at some critical points.
\begin{figure}[H]
\centering
\includegraphics[width=100mm]{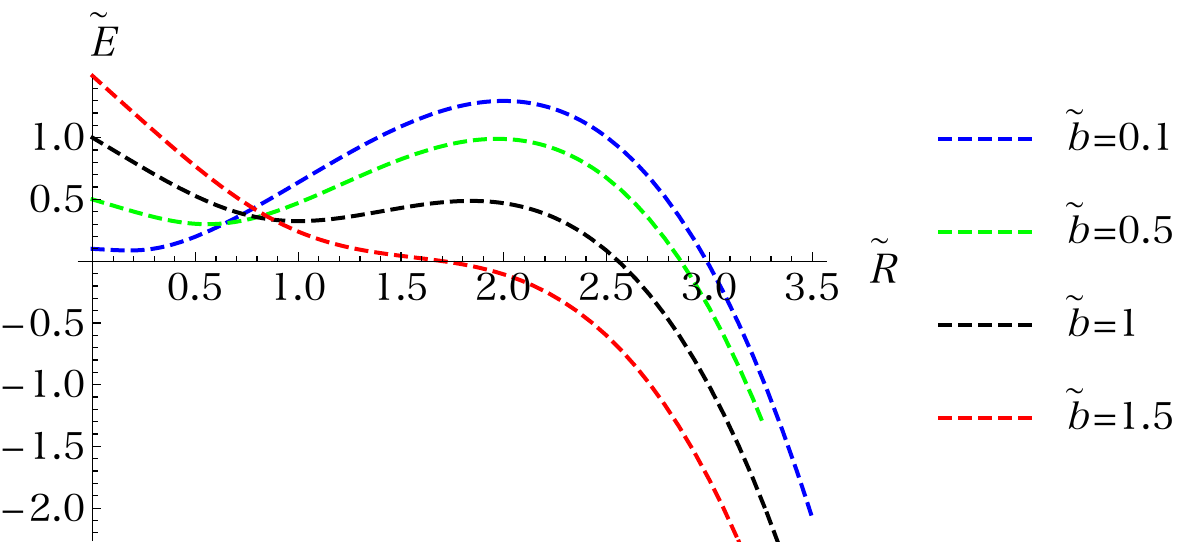}
\caption{Static energy for $\tilde{b}_{D3}=0.1,0.5,1,1.5$.}
\label{StaticE}
\end{figure}
\noindent
We can determine a critical strength of the magnetic field that the potential barrier disappears completely by studying the condition of the local minima. To find the coordinates, we need to solve the following equation
\begin{gather}
\frac{\partial\tilde{E}}{\partial\widetilde{R}}=\frac{2\widetilde{R}^3}{\sqrt{\widetilde{R}^4+\tilde{b}_{D3}^2}}-\sqrt{\widetilde{R}^4+\tilde{b}_{D3}^2}=0, \nonumber \\
\label{MaximalValue}
\Leftrightarrow\ \widetilde{R}^4-2\widetilde{R}^3+\tilde{b}_{D3}^2=0.
\end{gather}
Due to the numerical calculation, the range of magnetic field strength for the solutions to be real is given by
\begin{equation}
0\leq \tilde{b}_{D3} <\frac{3\sqrt{3}}{4} \simeq 1.299\cdots.
\end{equation}
When the field strength $\tilde{b}_{D3}$ is bigger than the critical value, $3\sqrt{3}/4$, there is no longer potential barrier. So the vacuum becomes unstable and decays without tunneling.   

It is necessary to find the bounce configuration that interpolates between the false and true vacua to obtain the decay rate of the system. The classical equation of motion is
\begin{equation}
    \label{EoM_bounce}
    \partial_{s}\left(\sqrt{\frac{\widetilde{R}^4+\tilde{b}_{D3}^2}{1+\dot{\widetilde{R}}^2}}-\tilde{b}_{D3}\widetilde{R}{}_{2}F_{1}\left(-\frac{1}{2}, \frac{1}{4}, \frac{5}{4}, -\frac{\widetilde{R}^4}{\tilde{b}^2_{D3}} \right) \right)=0.
\end{equation}
Solving the EOM for $d\widetilde{R}/ds$ yields
\begin{equation}
    \label{EOM_bounce2}
     \begin{aligned}
    &\therefore\frac{d\widetilde{R}}{ds}=\pm \frac{1}{\left[-C+\tilde{b}_{D3}\widetilde{R}{}_{2}F_{1}\left(-\frac{1}{2}, \frac{1}{4}, \frac{5}{4}, -\frac{\widetilde{R}^4}{\tilde{b}_{D3}^2} \right) \right]} \\
    &\hspace{5cm}\times\sqrt{\widetilde{R}^4+\tilde{b}^2-\left[-C+\tilde{b}_{D3}\widetilde{R}{}_{2}F_{1}\left(-\frac{1}{2}, \frac{1}{4}, \frac{5}{4}, -\frac{\widetilde{R}^4}{\tilde{b}_{D3}^2} \right) \right]^2}\ ,
    \end{aligned}
\end{equation}
where $C$ is a constant of integration, and the negative sign is assigned for later convenience.
Given the fact that a time derivative of the bounce solution must be zero at $\widetilde{R}=\widetilde{R}_{\mathrm{min}}$, an explicit form of $C$ yields
\begin{equation}
    \label{Constant}
    -C=\sqrt{\widetilde{R}^4_{\mathrm{min}}+\tilde{b}^2_{D3}}-\tilde{b}_{D3}\widetilde{R}_{\mathrm{min}}{}_{2}F_{1}\left(-\frac{1}{2}, \frac{1}{4}, \frac{5}{4}, -\frac{\widetilde{R}_{\mathrm{min}}^4}{\tilde{b}_{D3}^2} \right).
\end{equation}
Although it is difficult to find an analytical solution for the nonlinear equation, we can see the magnetic field dependence of the solution via numerical analysis (See Figure 3). From the Figure 3, we can read off the magnetic field dependence of the minimum radius of the bubble, $\widetilde{R}_{\mathrm{min}}=\widetilde{R}(s\rightarrow \pm \infty)$, and the maximum radius, $\widetilde{R}_{\mathrm{max}}=\widetilde{R}(s=0)$.
\begin{figure}[H]
\centering
\includegraphics[width=100mm]{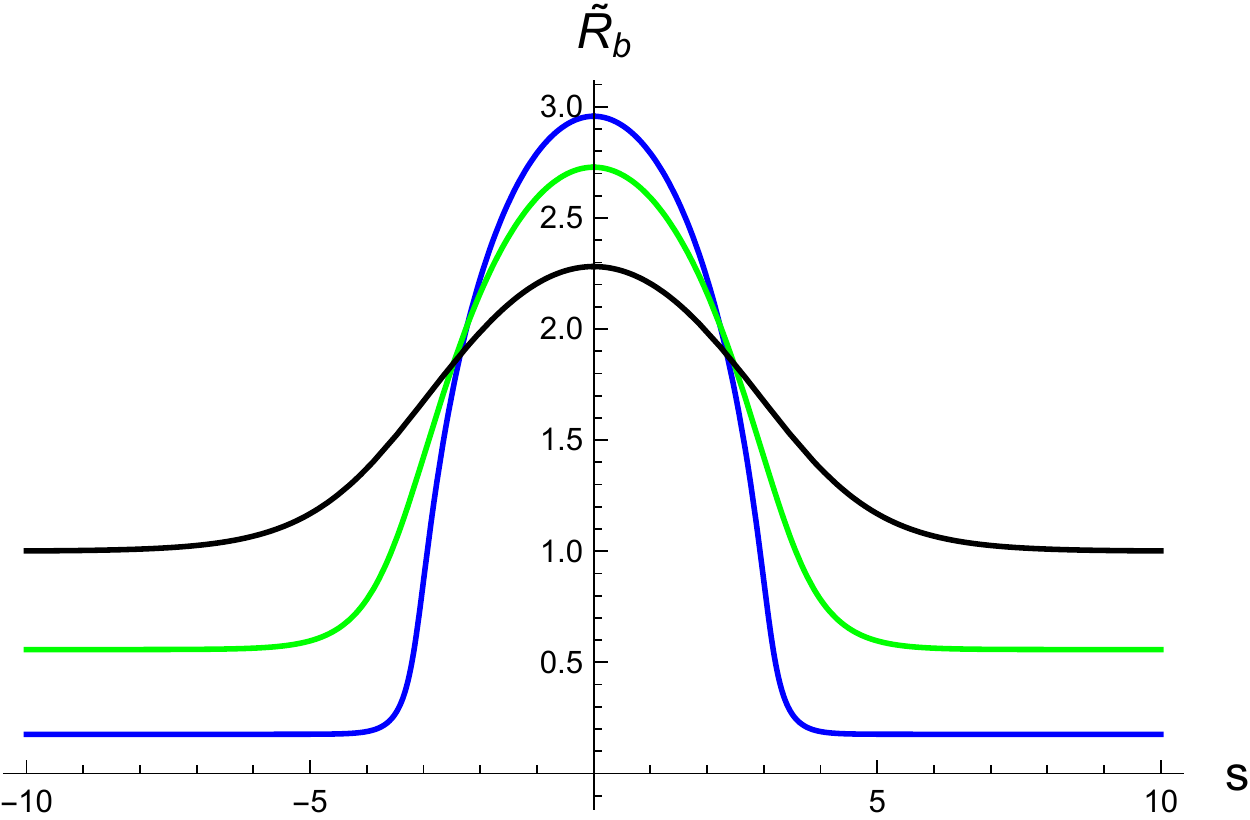}
\caption{Bounce solutions for $\tilde{b}_{D3}=0.1(\mbox{blue}),0.5(\mbox{green})\ \mbox{and}\ 1(\mbox{black})$.}
\label{DBIBounceSol.pdf}
\end{figure}
\noindent
As can be seen from Figure 3, when $b_{D3}$ is small, the bounce solution rises drastically near the origin and shows a non-standard shape. This property is due to the extremely small coefficients of the kinetic terms, which strongly indicate the contribution of nonlinearity. 
\begin{figure}[H]
\centering
\includegraphics[width=70mm]{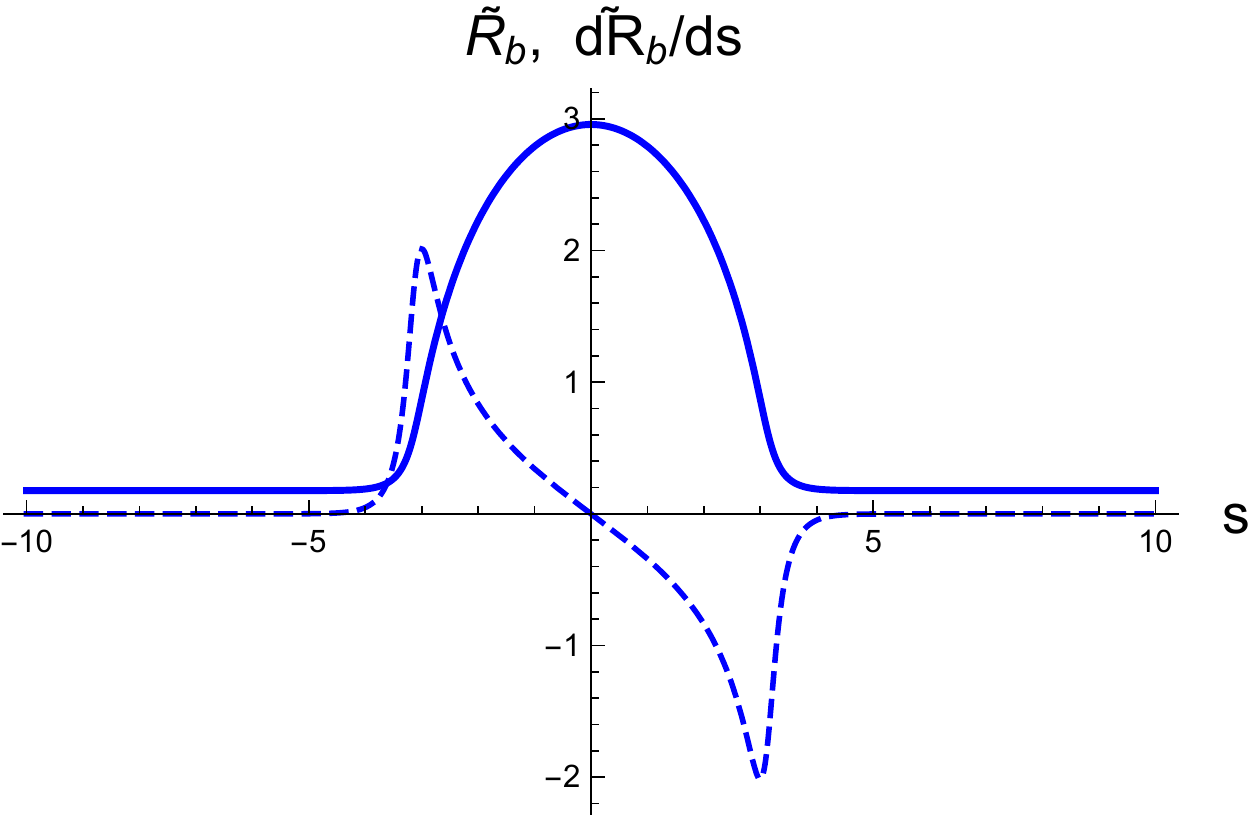}
\includegraphics[width=70mm]{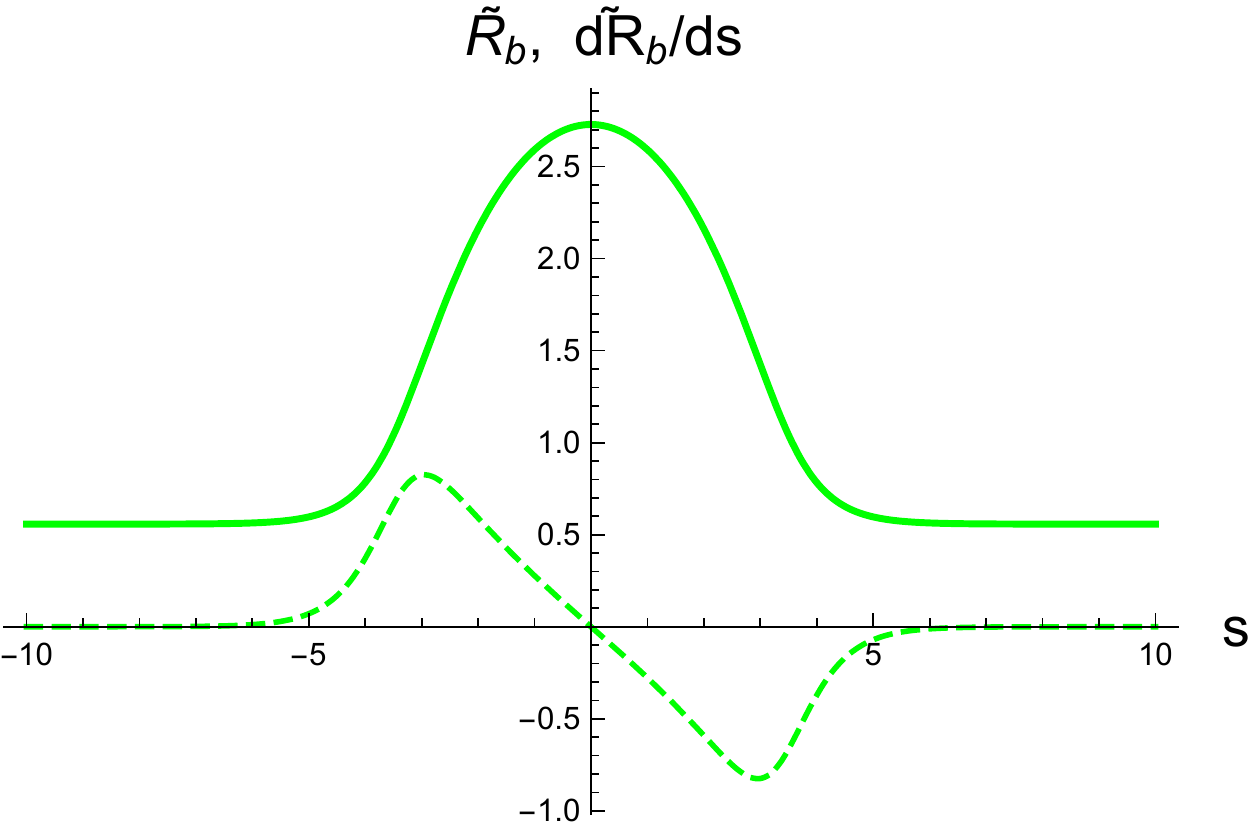}
\caption{Bounce solutions and their time derivatives for $\tilde{b}_{D3}=0.1 (\mbox{blue})\ \mbox{and}\ 0.5 (\mbox{green})$.}
\label{DBIBounceSolDif.pdf}
\end{figure}

Deducting the contribution of the trivial solution from the Euclidean action of the bounce solution, we get
\begin{equation}
\widetilde{B}=2\int^{\widetilde{R}_{\mathrm{max}}}_{\widetilde{R}_{\mathrm{min}}}d\widetilde{R}\sqrt{\widetilde{R}^4+\tilde{b}_{D3}^2-\left[-C+\tilde{b}_{D3}\widetilde{R}{}_{2}F_{1}\left(-\frac{1}{2}, \frac{1}{4}, \frac{5}{4}, -\frac{\widetilde{R}^4}{\tilde{b}_{D3}^2} \right) \right]^2}.
\label{BounceComputation}
\end{equation}
This bounce action depends on the magnetic field, and Figure 3 shows the numerical calculation for each magnetic field. The bounce action tends to decrease monotonically as the magnetic field increases. It implies that the catalyst accelerates vacuum decay in the parameter region where the exponential factor becomes dominant in calculating the life-time.
\begin{figure}[H]
\begin{center}
\includegraphics[width=70mm]{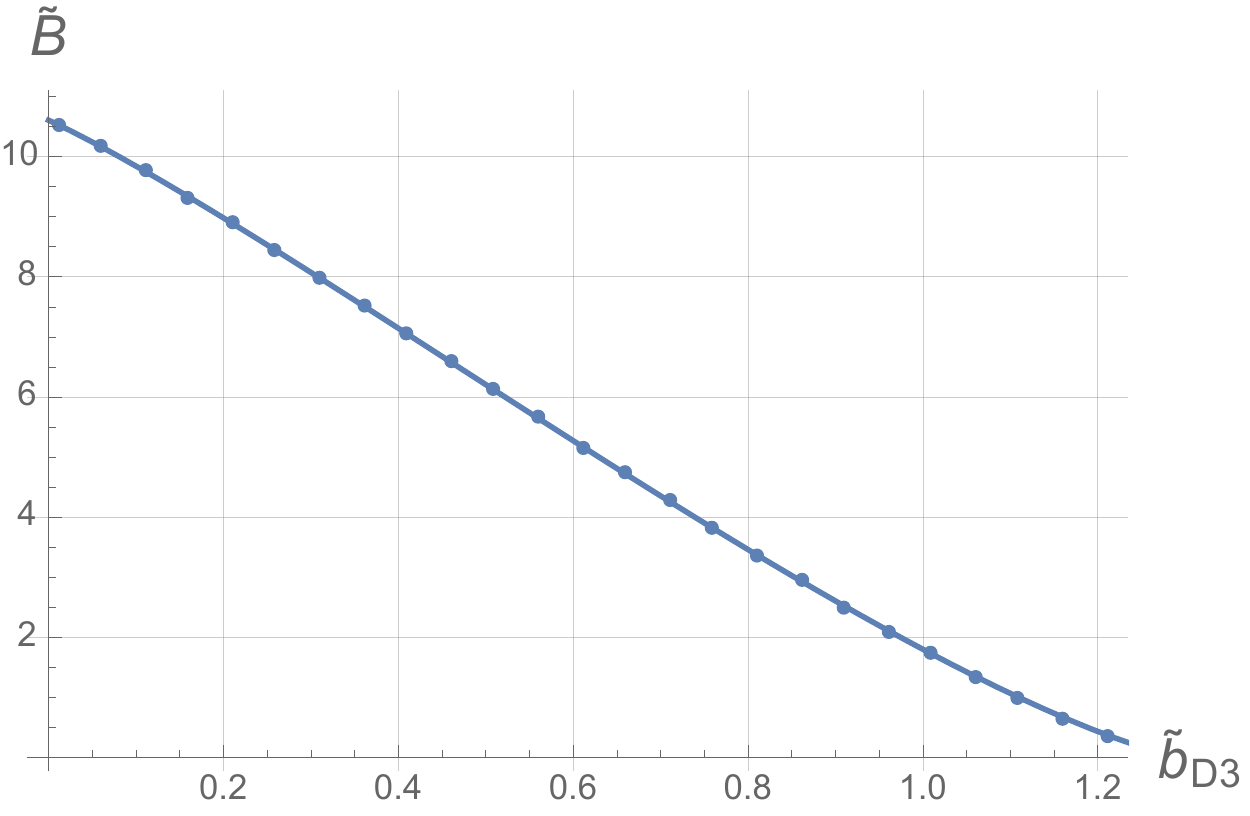}
\includegraphics[width=70mm]{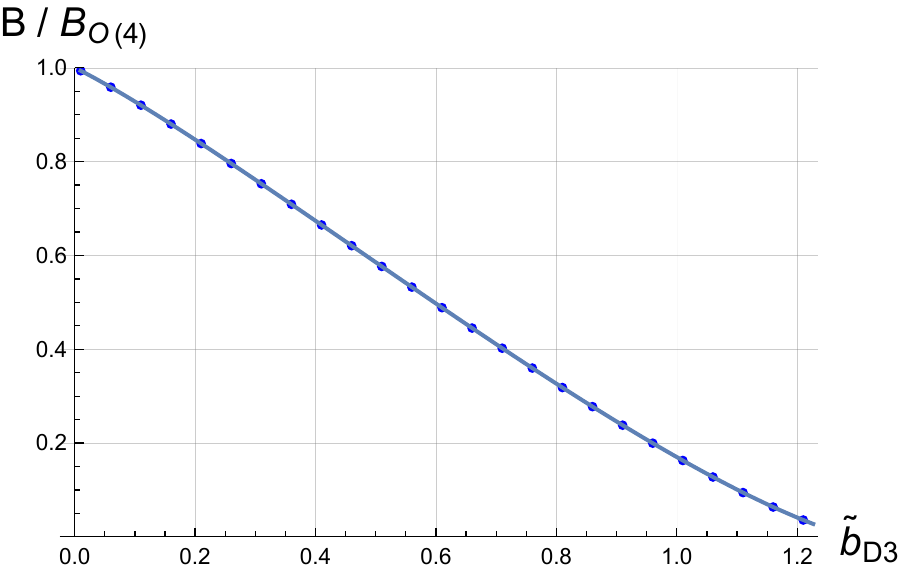}
\caption{The numerical calculation of the bounce action for each magnetic field. The right plot is normalized by the bounce action without catalyst \eqref{BO4}, and the ratio is less than one in all regions. }
\end{center}
\end{figure}

\subsection{Comments on zero mode normalization factor}
\label{Comments on zeromode normalization factor}

The zero mode can be written by 
\begin{equation}
\label{Nb}
\widetilde{R}_{0}=\frac{1}{\sqrt{N_{b}}}\frac{d\widetilde{R}_{b}}{ds},
\end{equation}
for any Lagrangian (see Appendix B for the brief proof).
We can compute a normalization factor via the following condition of the zero mode
\begin{align}
\label{prefactor}
N_{b}=\int^{\beta/2}_{-\beta/2}\left(\frac{d\widetilde{R}_b}{ds} \right)^2ds
=2\int^{\widetilde{R}_{\mathrm{max}}}_{\widetilde{R}_{\mathrm{min}}}d\widetilde{R}_{b} \frac{d\widetilde{R}_{b}}{ds}\ .
\end{align}
We emphasize here that  contrary to the standard argument by \cite{Coleman:1977py}, this normalization factor of the zero mode cannot be the same as the bounce action. This is essentialy comes from the non-linearity of the DBI action.  In fact, the normalization factor in the present model is given by 
\begin{eqnarray}
\label{normalization}
N_{b}&=&2\int^{\widetilde{R}_{\mathrm{max}}}_{\widetilde{R}_{\mathrm{min}}}d\widetilde{R}_{b} \frac{d\widetilde{R}_{b}}{ds} \nonumber \\
&=&2\int^{\widetilde{R}_{\mathrm{max}}}_{\widetilde{R}_{\mathrm{min}}}d\widetilde{R}_{b}\frac{\sqrt{\widetilde{R}_{b}^4+\tilde{b}_{D3}^2-\Big[-C+\tilde{b}_{D3}\widetilde{R}_{b}\cdot {}_2F_{1}\Big(-\frac{1}{2}, \frac{1}{4}, \frac{5}{4}, -\frac{\widetilde{R}_{b}^4}{\tilde{b}_{D3}^2} \Big) \Big]^2}}{-C+\tilde{b}_{D3}\widetilde{R}_{b}\cdot {}_{2}F_{1}\left(-\frac{1}{2}, \frac{1}{4}, \frac{5}{4}, -\frac{\widetilde{R}_{b}^4}{\tilde{b}_{D3}^2} \right)}\nonu.
\end{eqnarray}
There is an apparent discrepancy compared to the bounce action \eqref{BounceComputation}. We can also numerically estimate the magnetic field dependence of the normalization factor (see Figure 6). By comparing Figure 5 and Figure 6, we see an obvious difference between $N_b$ and $B$ in the intermediate region, although these show qualitatively similar behavior.  As a result, we need to replace the bounce action in the exponential pre-factor of the naive formula \eqref{NaiveFormula} with $N_b$.

\begin{figure}[H]
\begin{center}
\includegraphics[width=80mm]{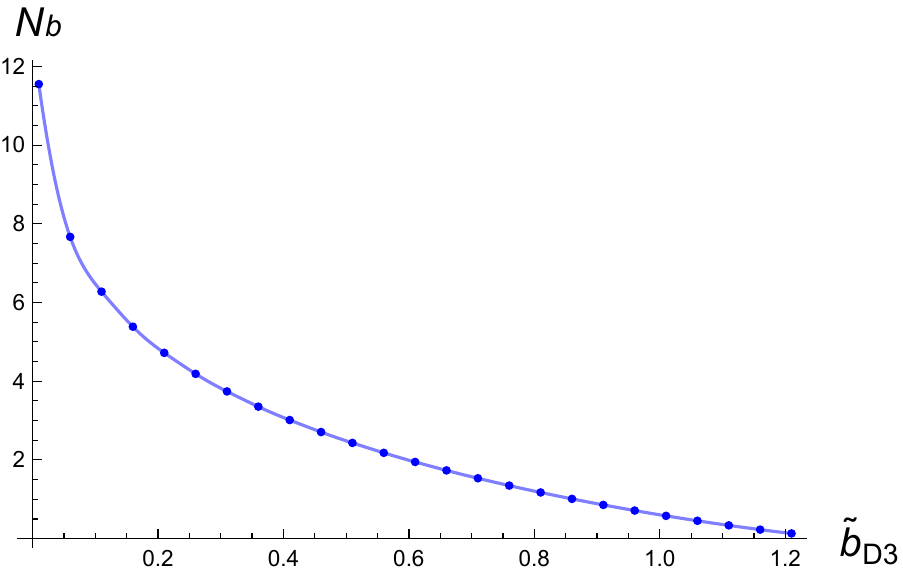}
\caption{The magnetic field dependence of the normalization constant.}
\end{center}
\end{figure}

Numerical results show that $N_{b}$ diverges at $\tilde{b}_{D3}=0$. This peculiar phenomenon means that the low-energy effective theory breaks down at this point. This zero magnetic field means the monopole no longer exists, and $O(4)$ symmetry is restored. It implies the existence of new zero modes, and the effective theory we consider, which does not include their contributions, is incorrect on this point.

\subsection{Life-time of vacuum by Euclidian path integral approach}
We can calculate the decay rate of the metastable state by the Euclidian path integral. The decay rate is given by
\begin{align}
    \label{DecayRate}
    \Gamma\equiv 2\mathrm{Im}E_{0}=\sqrt{\frac{N_{b}}{2\pi}}\left|\frac{\mathrm{det}^{\prime}\left[T[\widetilde{R}_{b}] \right]}{\mathrm{det}\left[T[\widetilde{R}_{\mathrm{min}}]\right]} \right|^{-1/2}e^{-B}\ .
\end{align}
Its inverse gives the life-time, $\tau=\Gamma^{-1}$. The determinant factor represents 1-loop quantum fluctuations around the classical solutions.
A general formula for the second variation of the Euclidean action is given by 
\begin{align}
\delta^2\widetilde{S}&=\int ds\left\{\frac{\partial^2 \widetilde{L}}{\partial \widetilde{R}^2}\left(\delta \widetilde{R} \right)^2+2\frac{\partial^2 \widetilde{L}}{\partial \widetilde{R} \partial \dot{\widetilde{R}}}\delta \widetilde{R} \delta \dot{\widetilde{R}}+\frac{\partial^2 \widetilde{L}}{\partial \dot{\widetilde{R}}^2}\left(\delta \dot{\widetilde{R}} \right)^2 \right\} \nonumber \\
&=\int ds \left\{-\left(\frac{d}{ds}\frac{\partial^2 \widetilde{L}}{\partial \dot{\widetilde{R}}^{2}} \left(\frac{d}{ds}\delta \widetilde{R} \right) \right)\delta \widetilde{R}+\frac{\partial^2 \widetilde{L}}{\partial \widetilde{R}^2}\left(\delta \widetilde{R} \right)^2-\frac{d}{ds}\frac{\partial^2 \widetilde{L}}{\partial \widetilde{R}\partial \dot{\widetilde{R}}}\left(\delta \widetilde{R} \right)^2 \right\}\ ,
\end{align}
where we used integration by parts in the second equality. 
We then define the Strum-Liouville operator over the interval $(-\beta/2,\beta/2)$,
\begin{align}
    T_{-\beta/2,\beta/2}
    \equiv -\frac{d}{ds}\left(\frac{\partial^2 \tilde{L}}{\partial \dot{\widetilde{R}}^{2}}\frac{d}{ds} \right)+\frac{\partial^2 \widetilde{L}}{\partial \widetilde{R}^2}-\frac{d}{ds}\frac{\partial^2 \widetilde{L}}{\partial \widetilde{R}\partial \dot{\widetilde{R}}}=-\frac{d}{ds}\left(P[\widetilde{R}]\frac{d}{ds} \right)+Q[\widetilde{R}]\ ,
    \label{SLO_def}
\end{align}
and the ratio of functional determinants is  
\begin{align}
    \label{det/det}
    \frac{\mathrm{det}^{\prime}\left[T[\widetilde{R}_{b}] \right]}{\mathrm{det}\left[T[\widetilde{R}_{\mathrm{min}}]\right]}
    =\frac{\mathrm{det}^{\prime}\left[-\frac{d}{ds}\left(P[\widetilde{R}_{b}]\frac{d}{ds} \right)+Q[\widetilde{R}_{b}] \right]}{\mathrm{det}\left[-\frac{d}{ds}\left(P[\widetilde{R}_{\mathrm{min}}]\frac{d}{ds} \right)+Q[\widetilde{R}_{\mathrm{min}}] \right]}\ .
\end{align}
Since the functional determinants suffer from UV divergence, we must regularize them appropriately. A well-established approach for this task is known for \textit{zeta function regularization} with a spectral zeta function, 
\begin{equation}
    \label{spectreZeta}
    \zeta(s;T_{-\beta/2,\beta/2})\equiv\sum_{j}\lambda^{-s}_{j}\ ,
\end{equation}
where $\lambda_{j}$ are eigenvalues of the Strum-Liouville operator defined above.
The zeta regularized determinant is then expressed by 
\begin{align}
    \label{zetaFD}
    \frac{\mathrm{det}^{\prime}\left[T[\widetilde{R}_{b}] \right]}{\mathrm{det}\left[T[\widetilde{R}_{\mathrm{min}}]\right]}
    =\exp\left[-\left(\zeta^{\prime}(0;T[\widetilde{R}_{b}])-\zeta^{\prime}(0;T[\widetilde{R}_{\mathrm{min}}]) \right) \right]\ ,
\end{align}
so all we need to do is to calculate the $s$ derivative of the spectral zeta function at $s=0$.

However, we soon face difficulties. As DBI Lagrangian \eqref{DBIlag_nondim} has a complicated kinetic term, the coefficient $P[\widetilde{R}_{b}]$ will be a time-dependent continuous function. We then have to calculate \eqref{zetaFD} for general Strum-Liouville operators, but this problem has been a challenging problem.
At this point, so-called \textit{contour deformation method} proposed by Kirsten and McKane is so powerful that we can do explicit computation \cite{Kirsten:2003py,Kirsten:2004qv}. According to their method, the derivative of the zeta function at $s=0$ is given by \cite{Fucci:2021gos}
\begin{align}
    \zeta^{\prime}(0;T_{A,B})
    &=i\pi n-\ln\left(2c\left|\frac{F_{m_{0}}}{\Gamma_{k_{0}}} \right| \right)\ .
\end{align}
See Appendix C for its derivation.
The natural number $n$ is the number of negative modes, which is equal to 1 in this case,
and the constant $c$ is determined by the weight function and $P[\widetilde{R}_b]$ in the Strum-Liouville operator.
$\Gamma_{k_{0}}$ is one of the expansion coefficients of the asymptotic series for the characteristic function in large $\lambda$ region, which is calculated as $\Gamma_{k_{0}}=\Gamma_{-1}=-2ic$ (see \cite{Fucci:2021gos} for more detailed discussions).
$F_{m_{0}}$ is the lowest order expansion coefficient for the small $\lambda$ expansion of the characteristic function in small $\lambda$ region, which is given by
\begin{equation}
    \label{F_expansion}
    F_{A,B}(\lambda)
    =F_{m_{0}}\lambda^{m_{0}}+\sum_{m=2}^{\infty}F_{m}\lambda^{m}\ ,
\end{equation}
where $m_{0}$ represents the multiplicity of the zero mode eigenvalue.
Then, the functional determinant from which zero mode is extracted is
\begin{align}
    \mathrm{det}^{\prime}\left[-\frac{d}{ds}\left(P[\widetilde{R}_{b}]\frac{d}{ds} \right)+Q[\widetilde{R}_{b}] \right]
    =-|F_{m_{0}}|\ .
\end{align}
In our case, the zero mode multiplicity equals one, so $F_{m_{0}}$ is the first-order coefficient of the asymptotic expansion.

When considering the expansion of the characteristic function, we need to know its original expression. We can write the characteristic function with two independent solutions of the Sturm-Liouville problem, $\psi^{1}_{\lambda}(s)$ and $\psi^{2}_{\lambda}(s)$,
\begin{align}
    \begin{aligned}
    F_{\varphi,R}(\lambda)
    &=\left\{\psi^{1}_{\lambda}\left(\frac{\beta}{2} \right)-\psi^{1}_{\lambda}\left(-\frac{\beta}{2} \right) \right\}\left\{\psi^{2[1]}_{\lambda}\left(\frac{\beta}{2} \right)-\psi^{2[1]}_{\lambda}\left(-\frac{\beta}{2} \right) \right\} \\
    &\hspace{1.5cm}-\left\{\psi_{\lambda}^{2}\left(\frac{\beta}{2} \right)-\psi_{\lambda}^{2}\left(-\frac{\beta}{2} \right) \right\}\left\{\psi^{1[1]}_{\lambda}\left(\frac{\beta}{2} \right)-\psi^{1[1]}_{\lambda}\left(-\frac{\beta}{2} \right) \right\}\ .
    \end{aligned}
\end{align}
by definition.
By use of the boundary conditions for $\psi^{1}_{\lambda}(s)$ and $\psi^{2}_{\lambda}(s)$,
\begin{equation}
    \psi^{1}_{\lambda}\left(-\frac{\beta}{2} \right)=\psi_{\lambda}^{2[1]}\left(-\frac{\beta}{2} \right)=1\ ,\quad \psi^{1[1]}_{\lambda}\left(-\frac{\beta}{2} \right)=\psi^{2}_{\lambda}\left(-\frac{\beta}{2} \right)=0\ ,
\end{equation}
one can simplify the expression of the characteristic function as
\begin{align}
    F_{\varphi,R}(\lambda)
    &=\left\{\psi^{1}_{\lambda}\left(\frac{\beta}{2} \right)-1 \right\}\left\{\psi^{2[1]}_{\lambda}\left(\frac{\beta}{2} \right)-1 \right\}
    -\psi^{2}_{\lambda}\left(\frac{\beta}{2} \right)\psi^{1[1]}_{\lambda}\left(\frac{\beta}{2} \right) \nonumber \\
    &=2-\psi^{1}_{\lambda}\left(\frac{\beta}{2} \right)-\psi^{2[1]}_{\lambda}\left(\frac{\beta}{2} \right)\ .
\end{align}
We can determine first-order coefficients of $\psi^{1}_{\lambda}(s)$ and $\psi^{2[1]}_{\lambda}(s)$ by Volterra integral equations
\begin{align}
    \psi^{1}_{\lambda,1}\left(s \right)
    &=\int^{s}_{-\beta/2}r(x)dxg\left(0,s,x \right)\psi^{1}_{0}(x) \nonumber \\
    &=\int^{s}_{-\beta/2}dx\left\{\psi^{1}_{0}\left(s \right)\psi^{2}_{0}\left(x \right)-\psi^{1}_{0}\left(x \right)\psi^{2}_{0}\left(s \right) \right\}\psi^{1}_{0}(x)\ , \\
    \psi^{2[1]}_{\lambda,1}\left( s\right)
    &=\int^{s}_{-\beta/2}r(x)dxg^{[1]}\left(0,s,x \right)\psi^{2}_{0}(x) \nonumber \\
    &=\int^{s}_{-\beta/2}dx\left\{\psi^{1[1]}_{0}\left(s \right)\psi^{2}_{0}\left(x \right)-\psi^{1}_{0}\left(x \right)\psi^{2[1]}_{0}\left(s \right) \right\}\psi^{2}_{0}(x)\ ,
\end{align}
where $g(\lambda,s,x)$ is Volterra Green's function.
Note that the weight function is fixed to one by the definition of the differential operator \eqref{SLO_def}.
Then we get 
\begin{align}
    \label{F_expansion2}
    \begin{aligned}
    {F_{\varphi,R,1}}
    &=-\int^{\beta/2}_{-\beta/2}dx\left[\left\{\psi^{1}_{0}\left(\frac{\beta}{2} \right)\psi^{2}_{0}\left(x \right)-\psi^{1}_{0}\left(x \right)\psi^{2}_{0}\left(\frac{\beta}{2} \right) \right\}\psi^{1}_{0}(x) \right. \\
    &\left.\hspace{3.5cm}+\left\{\psi^{1[1]}_{0}\left(\frac{\beta}{2} \right)\psi^{2}_{0}\left(x \right)-\psi^{1}_{0}\left(x \right)\psi^{2[1]}_{0}\left( \frac{\beta}{2}\right) \right\}\psi^{2}_{0}(x) \right]\ .
    \end{aligned}
\end{align}

To go a step further, we rewrite $\psi^{1}_{0}(s)$ and $\psi^{2}_{0}(s)$ as a linear combination of $\dot{\widetilde{R}}_{b}(s)$ and $\chi(s)$,
\begin{align}
    \left(
    \begin{array}{c}
    \psi^{1}_{0}(s) \\
    \psi^{2}_{0}(s) \\
    \end{array}
    \right)
    =\begin{pmatrix}
    a_{11} & a_{12} \\
    a_{21} & a_{22} \\
    \end{pmatrix}
    \left(
    \begin{array}{c}
    \dot{\widetilde{R}}_{b}(s) \\
    \chi(s) \\
    \end{array}
    \right)\ ,
\end{align}
where $\chi(s)$ is another zero mode solution independent of $\dot{\widetilde{R}}_{b}$ and defined by the energy derivative of the bounce solution.
The boundary conditions at $s=-\beta/2$ give
\begin{align}
    &\psi^{1}_{0}\left(-\frac{\beta}{2} \right)=a_{11}\dot{\widetilde{R}}_{b}\left(-\frac{\beta}{2} \right)+a_{12}\chi\left(-\frac{\beta}{2} \right)=1\ , \\
    &\psi^{2}_{0}\left(-\frac{\beta}{2} \right)=a_{21}\dot{\widetilde{R}}_{b}\left(-\frac{\beta}{2} \right)+a_{22}\chi\left(-\frac{\beta}{2} \right)=0\ , \\
    &{\psi^{1}_{0}}^{[1]}\left(-\frac{\beta}{2} \right)=P[\widetilde{R}_{b}\left(-\frac{\beta}{2} \right)]\left\{a_{11}\ddot{\widetilde{R}}_{b}\left(-\frac{\beta}{2} \right)+a_{12}\dot{\chi}\left(-\frac{\beta}{2} \right) \right\}=0\ , \\
    &{\psi^{2}_{0}}^{[1]}\left(-\frac{\beta}{2} \right)=P[\widetilde{R}_{b}\left(-\frac{\beta}{2} \right)]\left\{a_{21}\ddot{\widetilde{R}}_{b}\left(-\frac{\beta}{2} \right)+a_{22}\dot{\chi}\left(-\frac{\beta}{2} \right) \right\}=1\ .
\end{align}
Solving these equations, we find 
\begin{align}
    &(a_{11},a_{12},a_{21},a_{22}) \nonumber \\
    &\hspace{0.5cm}=\left(\frac{1}{\dot{\widetilde{R}}_{b}(-\beta/2)},0,-\frac{\chi(-\beta/2)}{P[\widetilde{R}_{b}(-\beta/2)]\dot{\widetilde{R}}_{b}(-\beta/2)\dot{\chi}(-\beta/2)},\frac{1}{P[\widetilde{R}_{b}(-\beta/2)]\dot{\chi}(-\beta/2)} \right)\ .
\end{align}
Hence the boundary values at $s=\beta/2$ are determined 
\begin{align}
    &\psi^{1}_{0}\left(\frac{\beta}{2} \right)=\frac{\dot{\widetilde{R}}_{b}(\beta/2)}{\dot{\widetilde{R}}_{b}(-\beta/2)}=1\ ,\hspace{2cm} 
    {\psi^{1}_{0}}^{[1]}\left(\frac{\beta}{2} \right)= 0\ , \\
    &\psi^{2}_{0}\left(\frac{\beta}{2} \right)=\frac{\chi(\beta/2)-\chi(-\beta/2)}{P[\widetilde{R}_{b}(-\beta/2)]\dot{\chi}(-\beta/2)}\ ,\hspace{2cm}
    \label{psi2beta/2}
    {\psi^{2}_{0}}^{[1]}\left(\frac{\beta}{2} \right)
w    =1\ ,
\end{align}
where we used the periodicity of the bounce solution and $\ddot{\widetilde{R}_{b}}(\pm \beta/2)=0$. Substituting these into \eqref{F_expansion2}, we obtain the following expression
\begin{align}
    F_{\varphi,R,1}&=\psi^{2}_{0}\left(\frac{\beta}{2} \right)\int^{\beta/2}_{-\beta/2}ds\left|\psi^{1}_{0}(s) \right|^{2} \nonumber \\
    &=\frac{\chi(\beta/2)-\chi(-\beta/2)}{P[\widetilde{R}_{b}(-\beta/2)]\dot{\chi}(-\beta/2)}\cdot \frac{1}{(\dot{\widetilde{R}}_{b}(-\beta/2))^{2}}\int^{\beta/2}_{-\beta/2}ds(\dot{\widetilde{R}}_{b}(s))^{2} \nonumber \\
    \label{F1st}
    &=\frac{N_{b}}{P[\widetilde{R}_{b}(-\beta/2)]\dot{\widetilde{R}}_{b}(-\beta/2)\dot{\chi}(-\beta/2)}\cdot \frac{\chi(\beta/2)-\chi(-\beta/2)}{\dot{\widetilde{R}}_{b}(-\beta/2)}\ .
\end{align}

To obtain a simpler form, let us focus on the periodicity of the bounce configuration
\begin{equation}
    \dot{\widetilde{R}{}}_{b}(t+\beta;C)=\dot{\widetilde{R}}_{b}(t;C)\ .
\end{equation}
Differentiating this by $\beta$ yields the following relation \cite{Marino:2015yie}
\begin{align}
    \frac{\chi(s+\beta/2)-\chi(s)}{\dot{\widetilde{R}}_{b}(s)}
    =-\left(\frac{\partial C}{\partial \beta} \right)^{-1}\ .
\end{align}

We can calculate the factor $P[\widetilde{R}_{b}(-\beta/2)]\dot{\widetilde{R}}_{b}(-\beta/2)\dot{\chi}(-\beta/2)$ in \eqref{F1st} by the energy conservation law, which is given by
\begin{equation}
    \left(\frac{d\widetilde{R}_{b}}{ds} \right)^{2}
    =\frac{{\widetilde{R}_{b}}^{4}+b^{2}}{\left[-C+b\widetilde{R}_{b}{}_{2}F_{1}\left(-\frac{1}{2}, \frac{1}{4}, \frac{5}{4}, -\frac{{\widetilde{R}_{b}}^{4}}{b^2} \right)\right]^{2}}-1\ .
\end{equation}
Differentiating by $C$ at $s=-\beta/2$ yields
\begin{align}
    2\dot{\widetilde{R}}_{b}\left(-\frac{\beta}{2} \right)\dot{\chi}\left(-\frac{\beta}{2} \right)
    &=\chi\left(-\frac{\beta}{2} \right)\frac{\partial}{\partial \widetilde{R}}\left.\left(\frac{{\widetilde{R}_{b}}^{4}+b^{2}}{\left[-C+b\widetilde{R}_{b}{}_{2}F_{1}\left(-\frac{1}{2}, \frac{1}{4}, \frac{5}{4}, -\frac{{\widetilde{R}_{b}}^{4}}{b^2} \right)\right]^{2}} \right)\right|_{s=-\beta/2} \nonumber \\
    &\hspace{1.5cm}+\frac{\partial}{\partial C}\left.\left(\frac{{\widetilde{R}_{b}}^{4}+b^{2}}{\left[-C+b\widetilde{R}_{b}{}_{2}F_{1}\left(-\frac{1}{2}, \frac{1}{4}, \frac{5}{4}, -\frac{{\widetilde{R}_{b}}^{4}}{b^2} \right)\right]^{2}} \right)\right|_{s=-\beta/2} \nonumber \\
    &=\frac{2}{\sqrt{{\widetilde{R}_{\mathrm{min}}}^{4}+b^{2}}}=\sqrt{\frac{2}{\widetilde{R}_{\mathrm{min}}^{3}}}\ .
\end{align}
One can relate this expression with $P_{\mathrm{min}}$, which is a boundary value of  $P[\widetilde{R}_{b}]$ at $s=-\beta/2$, 
\begin{align}
    P_{\mathrm{min}}=\left.\frac{\partial^{2}L_{E}}{\partial \dot{\widetilde{R}}_{b}^{2}} \right|_{s=-\beta/2}
    =\sqrt{2\widetilde{R}_{\mathrm{min}}^{3}} \ ,
\end{align}
then, 
\begin{align}
    \dot{\widetilde{R}}_{b}\left(-\frac{\beta}{2} \right)\dot{\chi}\left(-\frac{\beta}{2} \right)
    =\frac{1}{P_{\mathrm{min}}}
    \quad \Leftrightarrow \quad 
    P_{\mathrm{min}}\dot{\widetilde{R}}_{b}\left(-\frac{\beta}{2} \right)\dot{\chi}\left(-\frac{\beta}{2} \right)=1\ .
\end{align}
One can perform the same calculation for $Q_{\mathrm{min}}$.
Since we are interested in the value at $s=-\beta/2$, 
\begin{align}
    Q_{\mathrm{min}}&=\left.\frac{\partial^{2} L_{E}}{\partial \widetilde{R}_{b}^{2}} \right|_{s=-\beta/2} =\frac{1}{\sqrt{2\widetilde{R}_{\mathrm{min}}^{3}}}\left(6\widetilde{R}_{\mathrm{min}}^{2}-4\widetilde{R}_{\mathrm{min}}^{3} \right)\ ,
\end{align}
where we omitted the contribution from the second term in the definition.
Thus, \eqref{F1st} turns out to be 
\begin{align}
    F_{\varphi,R,1}=\frac{N_{b}}{P[\widetilde{R}_{b}(-\beta/2)]\dot{\widetilde{R}}_{b}(-\beta/2)\dot{\chi}(-\beta/2)}\cdot \frac{\chi(\beta/2)-\chi(-\beta/2)}{\dot{\widetilde{R}}_{b}(-\beta/2)}
    =-N_{b}\left(\frac{\partial C}{\partial \beta} \right)^{-1}\ .\label{Fm0}
\end{align}
This expression is identical to the one obtained when considering canonical theories \cite{Marino:2015yie}.

We must introduce a technical ingredient to calculate $\partial C/\partial \beta$. Following a procedure in \cite{Marino:2015yie}, let us rewrite $\beta$ as an integration by the bounce configuration
\begin{align}
    \label{beta_marino}
    \beta=2\int^{\widetilde{R}_{+}}_{\widetilde{R}_{-}}\left\{\frac{1}{\sqrt{({\dot{\widetilde{R}}_{b}}})^{2}}-\frac{1}{\sqrt{G[\widetilde{R}_{b}]}}+\frac{1}{\sqrt{G[\widetilde{R}_{b}]}} \right\}d\widetilde{R}_{b}\ .
\end{align}
$G[\widetilde{R}_{b}]$ is obtained as a Taylor expansion of $(d\widetilde{R}_{b}/ds)^{2}$ around $\widetilde{R}=\widetilde{R}_{\mathrm{min}}$ and $-C=E=\sqrt{\widetilde{R}_{\mathrm{min}}^{4}+b^{2}}-b\widetilde{R}_{\mathrm{min}}{}_{2}F_{1}\left(-\frac{1}{2}, \frac{1}{4}, \frac{5}{4}, -\frac{{\widetilde{R}}_{\mathrm{min}}^{4}}{b^2} \right)$, 
\begin{equation}
    G(\widetilde{R}_{b},C)\simeq \omega^{2}\left(\widetilde{R}-\widetilde{R}_{\mathrm{min}} \right)^{2}+\frac{2}{P_{\mathrm{min}}}(C+E)\ ,
\end{equation}
where $\omega=\sqrt{Q_{\mathrm{min}}/P_{\mathrm{min}}}$.
The third integral in \eqref{beta_marino} gives
\begin{align}
\begin{aligned}
    &2\int^{\widetilde{R}_{+}}_{\widetilde{R}_{-}}
    \frac{1}{\sqrt{G(\widetilde{R}_{b},C)}} d\widetilde{R}_{b} 
    =\omega^{-1}\log\left(\left(\widetilde{R}_{+}-\widetilde{R}_{\mathrm{min}}\right)^{2}\right)
    -\omega^{-1}\log \left(\frac{-(C+E)}{2Q_{\mathrm{min}}} \right)\ .
\end{aligned}
\end{align}
Then, the calculation for $\beta$ can be summarized as in
\begin{align}
    \beta&=2\int^{\widetilde{R}_{+}}_{\widetilde{R}_{-}}\left\{\frac{1}{\sqrt{(\dot{\widetilde{R}}_{b} )^{2}}}-\frac{1}{\sqrt{G(\widetilde{R}_{b},C)}} \right\}d\widetilde{R}_{b} \\
    &\hspace{3cm}+\omega^{-1}\log\left(\left(\widetilde{R}_{+}-\widetilde{R}_{\mathrm{min}}\right)^{2}\right)
    -\omega^{-1}\log \left(\frac{-(C+E)}{2Q_{\mathrm{min}}} \right) \nonumber \\
    &=2\int^{\widetilde{R}_{\mathrm{max}}}_{\widetilde{R}_{\mathrm{min}}}\left\{\frac{1}{\sqrt{(\dot{\widetilde{R}}_{b} )^{2}}}-\omega^{-1}\frac{1}{\widetilde{R}-\widetilde{R}_{\mathrm{min}}} \right\}d\widetilde{R}_{b} \nonumber \\
    &\hspace{3cm}+\omega^{-1}\log\left(\left(\widetilde{R}_{\mathrm{max}}-\widetilde{R}_{\mathrm{min}}\right)^{2}\right)
    -\omega^{-1}\log \left(\frac{-(C+E)}{2Q_{\mathrm{min}}} \right)\ ,
\end{align}
where we took a smooth limit at $C\rightarrow -E$ for the first integral.
As $\beta \rightarrow \infty$, solving the above equation for $C$ follows that 
\begin{align}
    C&=-2Q_{\mathrm{min}}\left(\widetilde{R}_{\mathrm{max}}-\widetilde{R}_{\mathrm{min}}\right)^{2} 
    \exp\left[2\omega\int^{\widetilde{R}_{\mathrm{max}}}_{\widetilde{R}_{\mathrm{min}}}\left\{\frac{1}{\sqrt{(\dot{\widetilde{R}}_{b} )^{2}}}-\omega^{-1}\frac{1}{\widetilde{R}-\widetilde{R}_{\mathrm{min}}} \right\}d\widetilde{R}_{b} \right] \nonumber \\
    &\hspace{10cm}\times\exp\left[-\omega\beta \right]-E\ .
\end{align}
$\beta$ derivative gives
\begin{align}
    \label{dC/dbeta}
    \begin{aligned}
        \frac{\partial C}{\partial \beta}
    &=2\left(\widetilde{R}_{\mathrm{max}}-\widetilde{R}_{\mathrm{min}}\right)^{2}\frac{Q_{\mathrm{min}}^{3/2}}{P_{\mathrm{min}}^{1/2}} \exp\left[2\omega\int^{\widetilde{R}_{\mathrm{max}}}_{\widetilde{R}_{\mathrm{min}}}\left\{\frac{1}{\sqrt{(\dot{\widetilde{R}}_{b} )^{2}}}-\omega^{-1}\frac{1}{\widetilde{R}-\widetilde{R}_{\mathrm{min}}} \right\}d\widetilde{R}_{b} \right] \\
    &\hspace{10cm}\times\exp\left[-\omega\beta \right]\ .
    \end{aligned}
\end{align}
Note that due to $\beta$ dependent factor, \eqref{Fm0} diverges exponentially in the limit of $\beta \rightarrow \infty$.
However, we can eliminate this divergence by regularization with the reference determinant $\mathrm{det}[T[\widetilde{R}_{\mathrm{min}}] ]$.
Therefore, we obtain a finite expression for the functional determinant, given by
\begin{align}
    &\frac{\mathrm{det}^{\prime}\left[-\frac{d}{ds}\left(P[\widetilde{R}_{b}]\frac{d}{ds} \right)+Q[\widetilde{R}_{b}] \right]}{\mathrm{det}\left[-\frac{d}{ds}\left(P[\widetilde{R}_{\mathrm{min}}]\frac{d}{ds} \right)+Q[\widetilde{R}_{\mathrm{min}}] \right]} \nonumber  \\
    \label{FD_final}
    &=-\frac{N_{b}}{2\left(\widetilde{R}_{\mathrm{max}}-\widetilde{R}_{\mathrm{min}} \right)^{2}}\frac{P_{\mathrm{min}}^{1/2}}{Q_{\mathrm{min}}^{3/2}}  \exp\left[-2\omega\int^{\widetilde{R}_{\mathrm{max}}}_{\widetilde{R}_{\mathrm{min}}}\left\{\frac{1}{\sqrt{(\dot{\widetilde{R}}_{b} )^{2}}}-\omega^{-1}\frac{1}{\widetilde{R}-\widetilde{R}_{\mathrm{min}}} \right\}d\widetilde{R}_{b} \right]\ .
\end{align}
We can calculate the square root of the absolute value, denoted $g(\tilde{b}_{D3})$, for each magnetic field strength numerically, and $g(\tilde{b}_{D3})$ is a monotonically increasing function, as shown in Figure 7.

As mentioned above, the life-time is defined as an inverse of the decay rate.
Plugging the calculation of the functional determinant into the definition \eqref{DecayRate}, we get
\begin{align}
    \tau
    =\frac{\sqrt{\pi}}{\left(\widetilde{R}_{\mathrm{max}}-\widetilde{R}_{\mathrm{min}} \right)}\frac{P_{\mathrm{min}}^{1/4}}{Q_{\mathrm{min}}^{3/4}} 
    \exp\left[-\omega\int^{\widetilde{R}_{\mathrm{max}}}_{\widetilde{R}_{\mathrm{min}}}\left\{\frac{1}{\sqrt{(\dot{\widetilde{R}}_{b} )^{2}}}-\omega^{-1}\frac{1}{\widetilde{R}-\widetilde{R}_{\mathrm{min}}} \right\}d\widetilde{R}_{b} \right]e^{B}\ .
\end{align}
As a result, the numerical calculation of the life-time for each magnetic field is shown in Figure 8.
\begin{figure}[H]
\begin{center}
 \includegraphics[width=80mm]{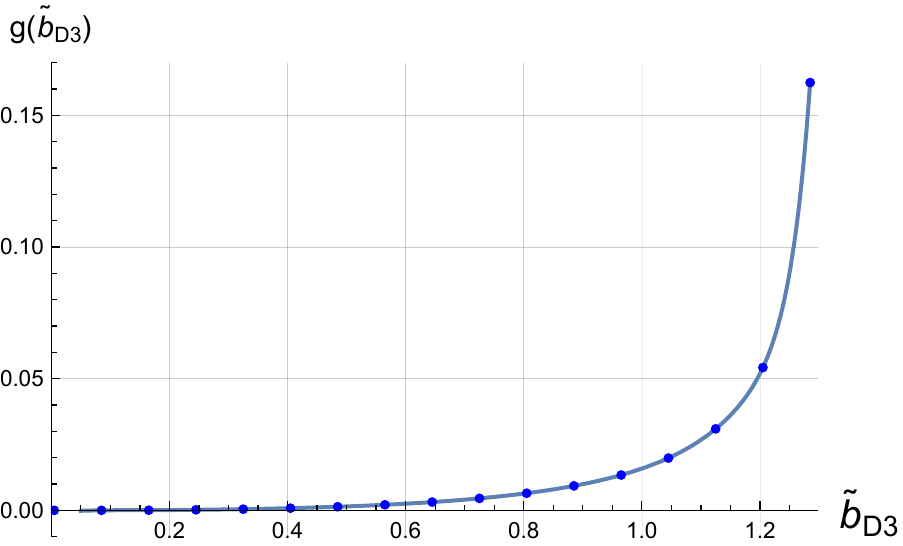}
\vspace{-.0cm}
\caption{\sl Magnetic field dependence of $g(\tilde{b}_{D3})$. We connected the points calculated for the arbitrary magnetic field strength smoothly.}
\end{center}
\end{figure}

\begin{figure}[H]
\begin{center}
\includegraphics[width=70mm]{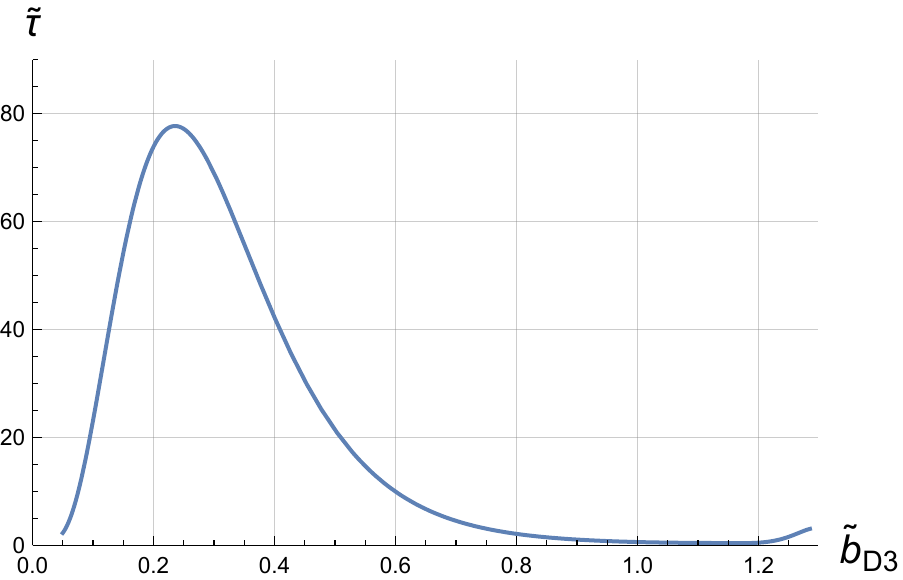}
\hspace{0.5cm}
\includegraphics[width=70mm]{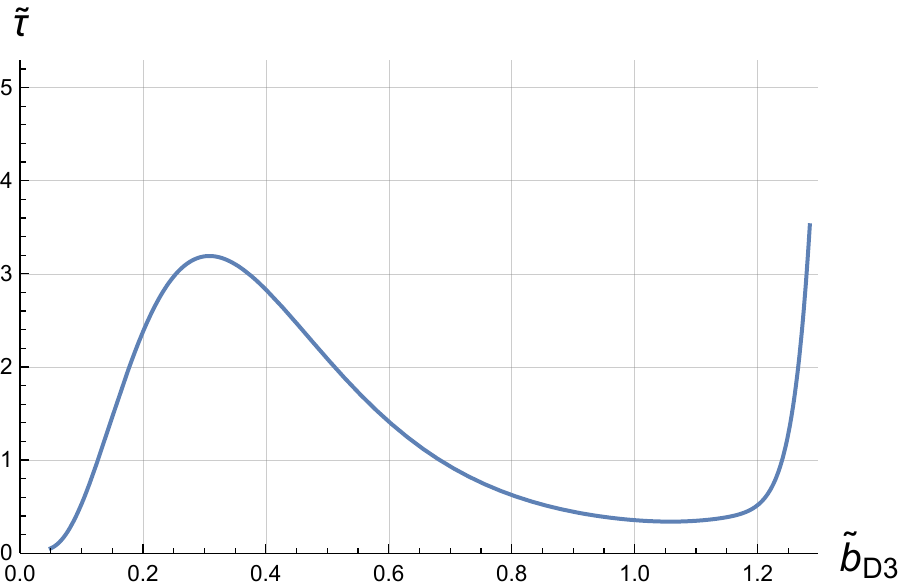}
\caption{The life-times for $A_B=1.5(\mbox{left}),1.1(\mbox{right})$. $\tilde{\tau}$ represents the dimensionless $\tau$ divided by $T_{DW}/\Delta V$.
In the small $\tilde{b}_{D3}$ region, we should note that the WKB approximation goes wrong, and its life-time shows strange behavior against naive expectations.
Considering this, we should exclude the region $\tilde{b}_{D3}\le 0.3 $ from our discussion.
}
\end{center}
\label{taucalc}
\end{figure}

\section{Higher order contributions and reduction to a cubic oscillator}
The approximations we used for our life-time calculation are twofold, the steepest descent approximation and the dilute gas approximation.
For the first one to be valid, the contribution of the classical solution must be sufficiently large, and the decrease of the exponential function around the saddle point must be sufficiently rapid. Thus, the condition for the approximation to be valid is that 
\begin{align}
A_B \widetilde{B}\gg 1\ ,\quad  A_B \det{}^\prime \big(P[R_b]{d^2\over ds^2}+Q[R_b] \big) \gg 1\ .
\end{align}
This condition is satisfied for sufficiently large $A_{B}$ even if the functional determinant is slightly smaller. However, when $A_B$ is not so large and the determinant is relatively small, this condition is violated. This phenomenon is unique to our noncanonical theory. The DBI action here is a low-energy effective theory, and the coefficients of the kinetic term at $b\to 0$ behave peculiarly. This tendency is one of the features when new particles appear at low energies, and in this case, the new zero modes would arise by the restoration of the $O(4)$ symmetry. We have not observed a similar phenomenon in the discussion of cubic oscillators such as \cite{Kleinert:1995ii} because the kinetic terms did not exhibit this specific behavior.

Another approximation is the dilute gas approximation. Let us assume a situation where an overall factor $A_B$ is enormous, and the semiclassical approximation is correct. Observing the instanton solution, the smaller $b_{D3}$, the sharper the change, and the less the instanton and anti-instanton overlap, shown in Figure 3. Therefore, the dilute gas approximation is correct when $b_{D3}$ is small. 

In summary, when $A_B\gg1$, both the saddle point method and the dilute gas approximation are correct from where $b_{D3}$ is small up to a certain magnitude. When the magnetic field becomes large enough, the instanton spreads out, the dilute gas approximation becomes incorrect, and the calculation fails, which is indicated by the divergence of $\tau$ near $b_{\rm crit}$. In the region where $b_{D3}$ is tiny, the value of the functional determinant becomes too small, which causes the saddle point approximation to fail. Indeed, the numerical calculation in Figure 8 shows that the life-time drops to zero in the small magnetic field region, contrary to intuition. We can observe the same tendency for $A_B\ge 1$. However, as mentioned above, $A_{B}$ is not so large that the second condition of the saddle point approximation is immediately violated. In other words, the reliable region of the calculation is narrower than in the case $A_B\gg1$.

As $b_{D3}$ increases, the life-time decreases monotonically at first but then blow-up near a critical point. It is physically unacceptable that the life-time increases while the potential stability worsens. This situation means that the leading contribution obtained by the WKB approximation breaks down. Thus, to perform the calculation accurately, it is necessary to calculate a contribution that includes the interaction between the bounce solutions, which was neglected in the dilute gas approximation. This procedure will, in principle, yield the correct result. However, there is a longstanding problem in this respect. Even if we perform the perturbative expansion in this way to obtain higher-order terms, the expansion series is known to be an asymptotic series with zero radii of convergence \cite{Dyson:1952tj}. A quartic oscillator is one of the well-known examples \cite{Bender:1969si,Bender:1973rz}. Therefore, obtaining accurate higher-order contributions is challenging in this model and difficult to study rigorously.

However, since we aim to evaluate the lifetime in a region with a strong catalytic effect, we do not need to take a detailed analysis for all $b_{D3}$.
If $b_{D3}$ is a nearly critical value, we can well approximate the potential by a cubic potential.
We, therefore, consider applying an analysis of cubic oscillators relying on variational perturbation expansion (VPE) \cite{Kleinert:1992tq,Kleinert:2004ev}.
The Lagrangian that Kleinert and Mustapic considered is given by \cite{Kleinert:1995ii}
\begin{align}
L_{KM}={m\over 2 }\dot{x}^2 + \Big[{m\omega^2 \over 2} x^2 -\lambda x^3 \Big]\ ,
\end{align}
where the overall sign of the potential is positive because we consider Euclidian theory. Let us transform the variable so that the potential minima are shifted from the origin, and the quadratic term vanishes, 
\begin{align}
x= y+{m\omega^2\over 6\lambda}
\end{align}
Then we obtain a rewritten Lagrangian
\begin{align}
L_{KM}={m\over 2 }\dot{y}^2 + \Big[{m^2 \omega^4\over 12\lambda }y  -\lambda y^3+{m^3\omega^6\over 108\lambda^2} \Big]+\cdots\ .
\end{align}

Let us expand our action around the inflection point.
Provided that the potential is almost flat near the critical point and the bounce solution shows an extremely moderate change in time, we can approximate the kinetic term as $\sqrt{1+\dot{R}^2}\simeq 1+\dot{R}^2/2$.
Expanding $R$ around the inflection point $R_v$ with $R=R_{v}+y$ and keeping it up to the third order, we obtain
\begin{align}
S=\int dt_E c^{-1} A_B \Big[ \sqrt{\widetilde{R}_{v}^4+\tilde{b}_{D3}^4} {\dot{y}^2\over 2} + V(\widetilde{R}_{v}, \tilde{b}_{D3})-f(\widetilde{R}_{v},\tilde{b}_{D3}) c^{-1}y  -h(\widetilde{R}_{v},\tilde{b}_{D3}) c^{-3} y^3 \Big]\ ,
\end{align}
\begin{align}
L \simeq  A_B \Big[ \sqrt{\widetilde{R}_{v}^4+\tilde{b}_{D3}^4} {\dot{y}^2\over 2} + V(\widetilde{R}_{v}, \tilde{b}_{D3})-f(\widetilde{R}_{v},\tilde{b}_{D3}) y  -h(\widetilde{R}_{v},\tilde{b}_{D3}) y^3 \Big]\ ,
\end{align}
where $y$ and $t_E$ are dimensioful values. Comparing this expression with the aforementioned $L_{KM}$, we obtain
\begin{eqnarray}
m=c^{-1}A_B\sqrt{\widetilde{R}_{v}^4+\tilde{b}_{D3}^2}\ , \quad  {m^2 \omega^4\over 12\lambda }=c^{-2}A_Bf(\widetilde{R}_{v},\tilde{b}_{D3}) \ , \quad \lambda =c^{-4}A_Bh (\widetilde{R}_{v},\tilde{b}_{D3})\ .
\end{eqnarray}

With the above comparison in mind, we can use the calculation result of complex energy for a cubic potential, shown in \cite{Kleinert:1995ii}.
\begin{figure}[H]
\begin{center}
\includegraphics[width=80mm]{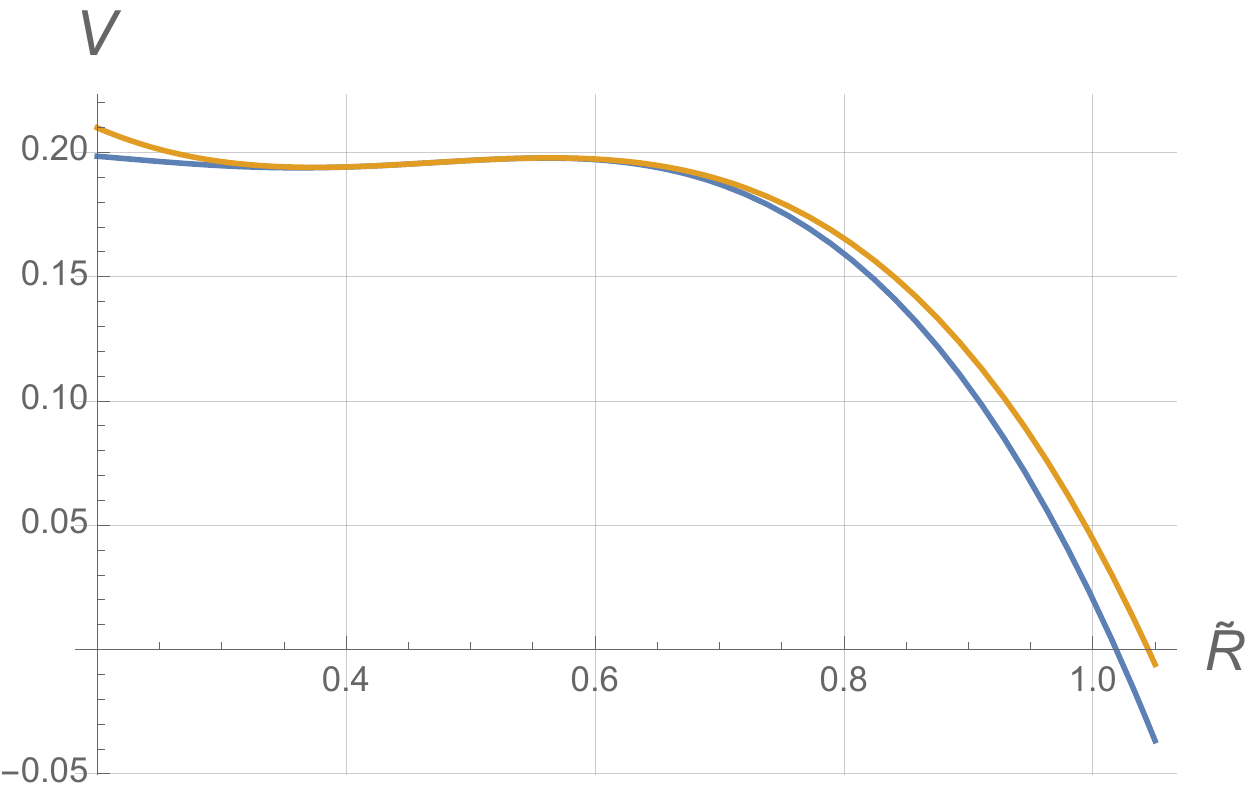}
\caption{The blue line represents our DBI potential, and the orange one is the fitting curve with cubic potential.
}
\end{center}
\end{figure}

Let us consider the limit where the potential barrier vanishes in the case of the previous cubic oscillators.
Kleinert and Mustapic have derived the decay rate up to a leading order by using VPE, which is given by
\begin{align}
\label{KM_decay}
\Gamma =-2 {\rm Im} E_0\simeq 2\times (0.448) \left({\lambda^2 \over m^3} \right)^{1/5}\left( 1-0.186 \Big({m^3 \omega^5\over \lambda^2} \Big)^{2\over 5} \right) \ .
\end{align}
The number $-0.186$ was read from the graph in \cite{Kleinert:1995ii} and may include a discrepancy of about factor 2, but that does not significantly change the argument. Figure 10 shows the lifetime calculation for $b_{D3}$ near the critical value. It shows that the lifetime becomes shorter as the instability increases.

\begin{figure}[H]
\begin{center}
\includegraphics[width=80mm]{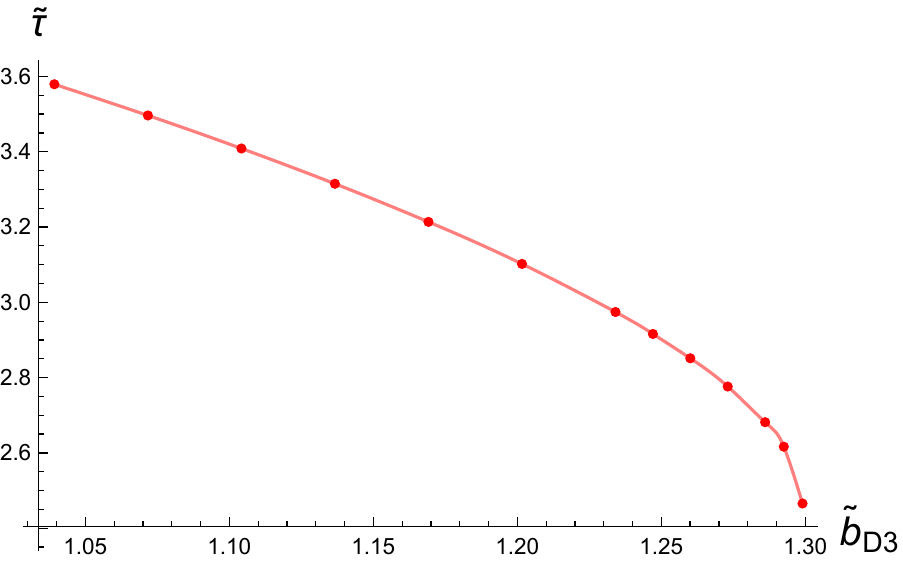}
\caption{A dimensionless lifetime, where we set $A_B=1$.
}
\end{center}
\end{figure}

Using the above limits, we can roughly evaluate the magnetic field dependence of the life-time by connecting the results of WKB calculation when $b_{D3}$ is small with those around $b_{\rm cr}$. This treatment includes the assumption that the life-times will monotonically decrease as the potential loses stability. Interestingly, the decay rate does not go to zero even at the limit where the potential barrier disappears. Therefore, let us compare the TCC condition with this limit of the life-time, $\tau_{\rm cr}$.
We can calculate the limit as
\begin{eqnarray}
\tau_{\rm cr}\simeq {1\over 2\times 0.448} {T_{DW}\over \Delta V}A_B^{1\over 5}  \left({h^2\over (\widetilde{R}_v^4+\tilde{b}_{D3}^2)^{3/2}} \right)^{-1/5}\Big|_{cr}\simeq 2.46 A_B^{1\over 5}{T_{DW}\over \Delta V}\ ,
\end{eqnarray}
where we used \eqref{KM_decay}.
Factor ${T_{DW}\over \Delta V}$ emerges by rewriting dimensionless time into a dimensioful quantity.
Therefore, the TCC condition is given by
\begin{eqnarray}
2.46 A_B^{1\over 5}{T_{DW}\over \Delta V} \le H_I^{-1} \log {M_{\rm pl}\over H_I}\ ,
\end{eqnarray}
where $H_I=\sqrt{V_{\rm meta}\over M^2_{\rm pl}}$, and $M_{\rm pl}=\sqrt{3\over 8\pi G}$ is the reduced Planck mass.
We assume $\Delta V=V_{\rm meta}-V_{\rm true}=n V_{\rm meta}$.
Since this is a stringy metastable state, the energy density of the vacuum is assumed to be $V_{\rm meta}=M_{\rm st}^4$. We also assume $T_{DW}=M_{\rm st}^3/g_s$ for a domain wall tension, where $g_s$ is the string coupling constant that is expected to satisfy $g_s<1$.
We define
\begin{align}
F={2.46\over 2} \Big( {4\pi\over n^8} \Big)^{1\over 5} \Big({M_{\rm st}\over M_{\rm pl}}\Big) \Big( \log {M_{\rm pl}\over M_{\rm st}} \Big)^{-1}\ ,
\end{align}
then rewrite the swampland condition as $\log F\le {9\over 5}\log g_s$. As shown in Figure 11, if $M_{\rm st}$ is about the same as the Planck scale, the TCC condition cannot be satisfied even if we account for catalytic effects. Bearing that the string coupling constant $g_s$ must be small for the perturbation to be a good approximation, the string scale must be small compared to $M_{\rm pl}$. However, interestingly, if it is reduced by about one order of magnitude, the condition is satisfied to some extent due to rapid damping near the origin. 

\begin{figure}[H]
\begin{center}
\includegraphics[width=100mm]{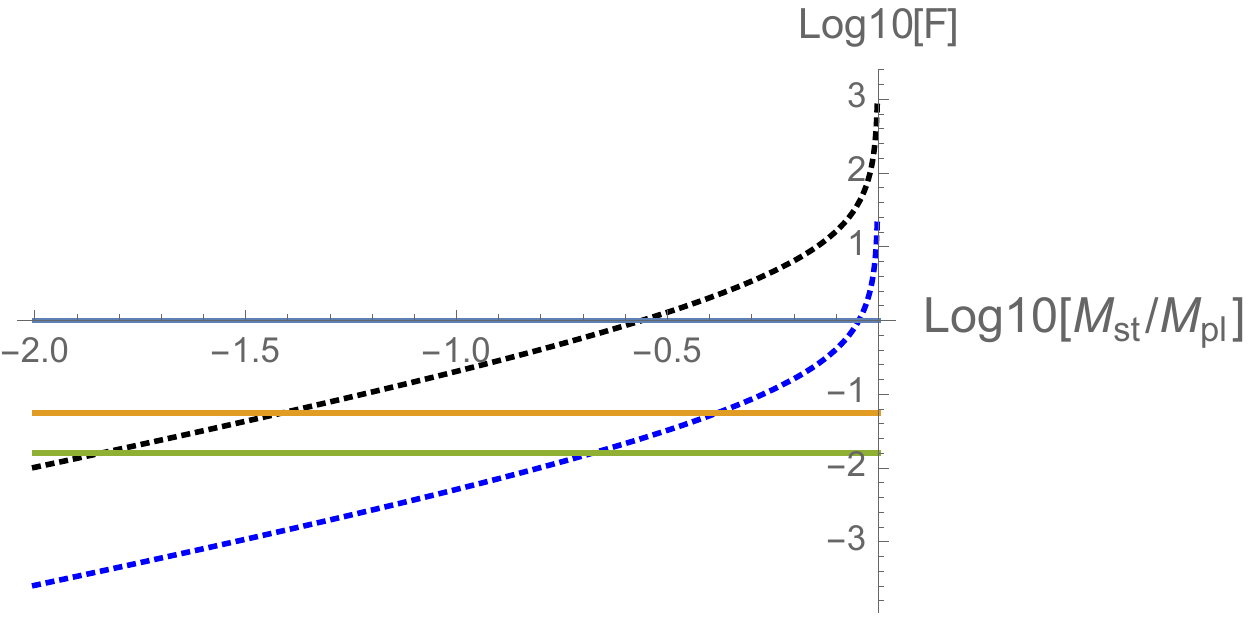}
\caption{Each black and blue dashed line represents $n=1,10$, and each blue, orange and green solid line represents $g_s=1$, $1/5$ $1/10$.
}
\end{center}
\end{figure}

\begin{figure}[H]
\begin{center}
\includegraphics[width=80mm]{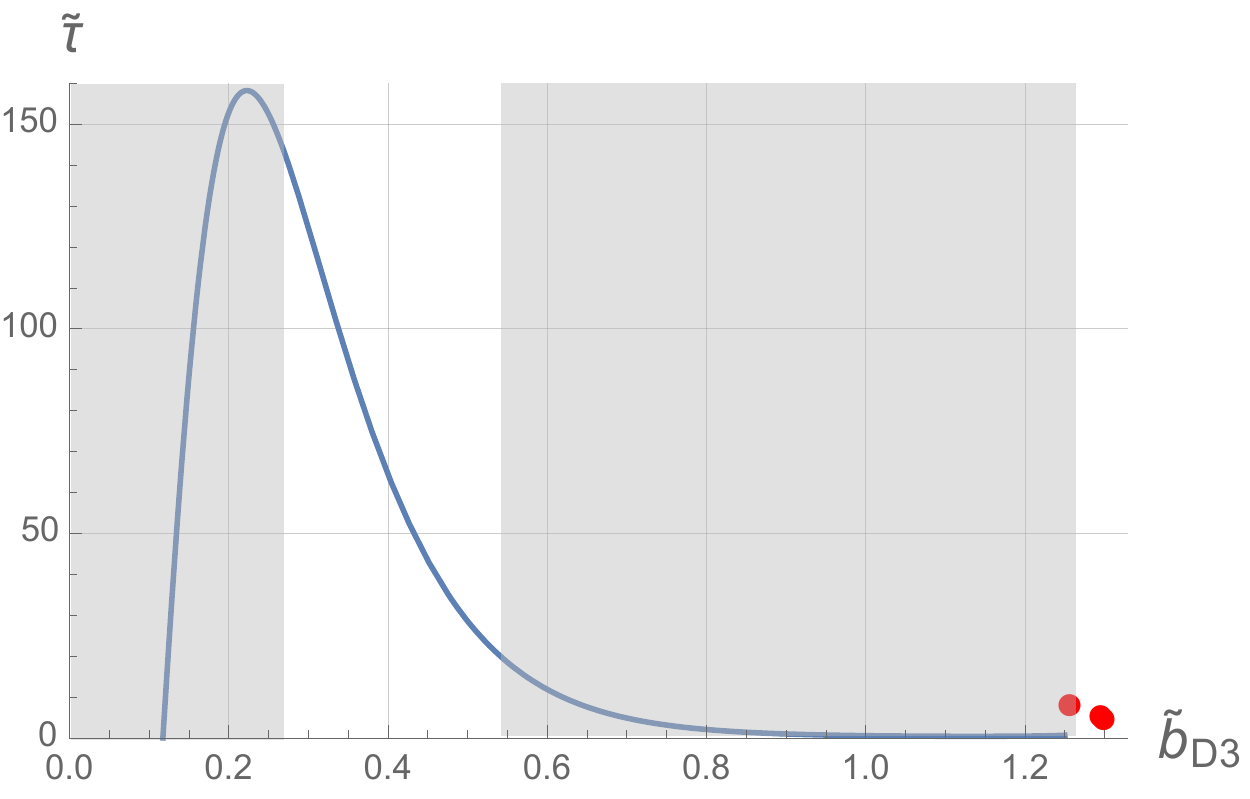}
\caption{The result of the WKB approximation and the VPE for $A_B=1.5$. The former is estimated to have a slightly shorter lifetime near the critical value, which is consistent with an argument beyond the dilute gas approximation. Although neither calculation is reliable in the orange region, based on physical assumptions, we expect a monotonically decreasing function connecting the two.
}
\end{center}
\end{figure}

Beyond the dilute gas approximation, ${\rm Im}E_{0}$ tends to be smaller when we consider the interaction between bounce solutions, according to Kleinert and Mustapic \cite{Kleinert:1995ii} . Therefore, its inverse, the lifetime, becomes larger. The results shown in Figure 12 are consistent with this statement since the life-time is estimated to be shorter when we ignore the interaction.

\section{Summary}

In this paper, we first reviewed the trans-Planckian censorship conjecture, which has recently drawn much attention in light of the swampland program in string theories. We have applied the conjecture to a metastable vacuum in string theory under an assumption of positive energy of the vacuum. A monopole like D3-brane can play a role of a catalyst that induces catalytic decay of the de Sitter vacua and make the life-time of the vacua much shorter. Considering the decay, we have searched for the allowed parameter spaces where the lifetime of vacua does not conflict with the conjecture.

It would be interesting to explore further by using known solutions in various dimensions and study allowed region of the swampland program. Throughout this paper, we took the same assumption imposed in \cite{TCC1} and assumed the existence of the de Sitter vacua. However, the strongest version of the de Sitter conjecture \cite{SW5} does not allow them. So we would like to discuss the possibility of the existence of a de Sitter vacuum in the presence of any catalyst in the context of string theory. Clearly, it is beyond the scope of this paper, so we will leave it for future work. 

\section*{Acknowledgments}
We are very grateful to Yutaka Ookouchi for providing us with this project and the collaboration in the early stages. We also thank Issei Koga and Yuri Okubo for their valuable discussions. This work is supported by the Kyushu University Leading Human Resources Development Fellowship Program.

\appendix

\section{Derivation of action}
In the main text, we provided the explicit example of semi-classical vacuum decay induced by impurities and discussed TCC bound on the parameter spaces where the gravitational effect is sub-dominant. D3-branes, which play the role of catalyst, dissolved into (anti) D5-branes and constructed the bound state in our scenario. In this Appendix, we first review a geometric construction of the metastable state using D5-branes and anti D5-branes, then show an explicit expression of the Lagrangian, which describes the bound state of the bubble and impurity.

\subsection{Quick review of geometrically induced vacua}

We will explain the setup for our model and give the derivation of the Lagrangian.
While this model has already been discussed in \cite{Kasai:2015exa}, we will show a refinement for the more accurate description of a small bubble\footnote{Although we have assumed that the gain of the energy from the true vacuum was proportional to $R^{3}$, this is incorrect when the bubble is tiny.}.

We consider a metastable vacuum because we would like to examine a scenario that constructs dS vacua by enhancing the instability with branes as catalysis, even if a swampland conjecture naively rejects the possibility in string theory. We adopt an already-known construction method here, which is based on a special geometry including singular points.
This method is appropriate for our purpose because the behavior of the Dp-brane around the singular points domains the decay process. 
The effective theory on the Dp-branes is governed by the string length, $l_{\rm st}$.
Brane configuration engineered near a singularity creates a mass hierarchy. Given the scale of compactification is $V$, $V/G_{10}$ give a Planck scale in 4 dimensions, and we can take the Newton constant large enough if we hold $G_{10}\simeq l_{\rm pl}^6$.
This yields a brane limit such that the 4D gravity contribution is negligible, but the string correction is not.

First, we explain an internal space and its properties.
\begin{equation}
\label{InternalSpace}
0=z_1^2+z_2^2+z_3^2 +W^{\prime}(z_4)^2~, \qquad W'(z_4) = g (z_4-a_1)(z_4-a_2),
\end{equation}
where $z_{i}$ are complex variables, $z_i \in \mathbb{C}$.
We make the geometric variables dimensionless via normalization by $l_{\rm st}$ to consider the brane limit, which $l_{\rm st}$ is finite.
This geometry has two special cycles, named $[C_1]$ and $[C_2]$.
Interestingly, these cycles are not independent, that is, $[C_1]+[C_2]=0$.
However, The transition of a Dp-brane wrapped around $[C_1]$ to $[C_2]$ requires extra energy because it has to go through a larger $S^2$ between them, and vice versa. Thus, we can construct the metastable states by wrapping Dp-branes and anti Dp-branes around each cycle.

When considering brane dynamics, We often assume that Dp-branes wrap a nontrivial cycle. Here, we briefly review the derivation of the induced metric to calculate an effective action on the Dp-brane, the DBI action. The induced metric is defined as
\begin{align}
g_{\alpha \beta}= g_{\mu \nu} {\pa X^\mu \over \pa \sigma^\alpha} {\pa X^\nu \over \pa \sigma^\beta}\ ,
\end{align}
where $X^\mu$ is an embedding function, and $\sigma^\alpha$ is spacial coordinates on Dp-branes.
In string theory, nontrivial cycles occur in the internal space, so let us also assume $x^{4,5\cdots }$ as the space coordinates. We will first treat the relatively simple case of $ \mathbb{S}^2$ since it will be helpful in the following discussion. $ \mathbb{S}^2$ of radius $L$ in $x^{4,5,6}$ space is a set of points satisfying the following conditions
\begin{eqnarray}
(x^4)^2+(x^5)^2+(x^6)^2=L^2\ .
\end{eqnarray}
Thus, we use the following parameter equation
\begin{eqnarray}
 X^4 =L \sin \theta_I \cos \varphi_I \ , \quad X^5= L\sin \theta_I \sin \varphi_I\ , \quad X^6=L\cos \theta_I \nonu\ ,
\end{eqnarray}
as the embedding functions.
Let us use these relations to find each component of the induced metric.
Since $a$ is a coordinate of the internal space,
\begin{eqnarray}
&&\pa_\theta X^a \pa_\theta X_a = L^2\cos^2 \theta_I \cos^2 \varphi_I +L^2 \cos^2 \theta_I \sin^2 \varphi_I +L^2 \sin^2 \theta_I =L^2\ ,\nonu \\
&&\pa_\theta X^a \pa_\varphi X_a = -L^2\cos \theta_I \sin \theta_I \cos \varphi_I \sin \varphi_I +L^2 \cos \theta_I \sin \theta_I \sin \varphi_I \cos \varphi_I =0\ , \nonu \\
&&\pa_\varphi X^a \pa_\varphi X_a =L^2 \sin^2 \theta_I\ , \nonu 
\end{eqnarray}
where the subscript ``${}_I$'' means the internal space. Then, we find the induced metric
\begin{eqnarray}
g_{\alpha \beta} =  \begin{pmatrix} 
 L^2  & 0 \\ 
 0 & L^2 \sin^2 \theta_I \\ 
 \end{pmatrix}
 \ .
\end{eqnarray}

Next, we consider a slightly more complicated situation in which the internal space's geometry is given by $ \mathbb{S}^3$. Since all we have to do is to add one more dimension to the above discussion, the constraint condition that must be satisfied is given by
\begin{align}
(x^4)^2+(x^5)^2+(x^6)^2+(x^7)^2=L^2\ .
\end{align}
It would be useful to consider the geometry as a set of $ \mathbb{S}^2$ with different radii. The radii of the $ \mathbb{S}^2$ are different at every $x^7$ location. The radius of $ \mathbb{S}^2$ corresponding to an angle $\psi_I$ is given as $L \sin \psi_I$, then the embedding function are 
\begin{eqnarray}
X^4=L \sin \psi_I \sin \theta_I \cos \varphi_I  \ , \ X^5=L \sin \psi_I \sin \theta_I \sin \varphi_I \ , \ X^6=L \sin \psi_I \cos \theta_I  \ , \ X^7= L \cos \psi_I\ . \nonu
\end{eqnarray}
Therefore, we obtain the induced metric as
\begin{align}
g_{\alpha \beta} =  \begin{pmatrix} .
 L^2&0  & 0 \\ 
0&  L^2 \sin^2 \psi_I & 0\\ 
0&  0 & L^2 \sin^2 \psi_I \sin^2 \theta_I \\ 
 \end{pmatrix}
 \ ,
\end{align}
where we write in the order $\psi_I, \theta_I ,\varphi_I$.

\subsection{The total energy of the system}

Now that we are ready, we will consider the situation where we wrap D5-brane and anti D5-brane around $ \mathbb{S}^2$ and also wrap D3-branes as catalysts. Consider the DBI action when a part of the D5-brane is wrapped around $ \mathbb{S}^2$ in $ \mathbb{S}^3$. As for the embedding function, 
\begin{eqnarray}
&&X^\mu =x^\mu \ , \ (\mu=0\cdots 3)\ , \quad X^i= \mbox{const.\ , \  (\mbox{the other directions})}  \nonu \\
&&X^4=L \sin \psi_I \sin \theta_I \cos \varphi_I  \ , \ X^5=L \sin \psi_I \sin \theta_I \sin \varphi_I \ , \ X^6=L \sin \psi_I \cos \theta_I  \ , \ X^7= L \cos \psi_I\ . \nonu
\end{eqnarray}
We will use this to consider the DBI action and will include $B^{NS}_{\mu \nu}$ in the $(\theta_I , \varphi_I)$ direction as a background field. Let us denote it as $B_{\theta \varphi}^{NS}$. Also, note that the direction in which the D5-brane wraps around is $(\theta_I, \varphi_I)$ and not in the $\psi_I$ direction. Then, 
\begin{eqnarray}
\det \Big( \pa_{\sigma_1} X^\mu \pa_{\sigma_2} X_\mu \Big) &=& \det  \begin{pmatrix} 
 \eta_{\mu \nu} &  0 &0 \\ 
   0 & L^2 \sin^2 \psi_I  &B_{\theta \varphi}^{NS}\\ 
  0& -B_{\theta \varphi}^{NS} & L^2 \sin^2 \psi_I \sin^2 \theta_I \\ 
 \end{pmatrix} \nonu \\
 &=& -  \Big( L^4 \sin^4\psi_I \sin^2 \theta_I +(B^{NS}_{\theta \varphi})^2 \Big)\ .
 \label{DBIcal1}
\end{eqnarray}
Substituting this into the DBI action yields
\begin{align}
S_{D5} =-T_{D5} \int d^4x d\theta_I d\varphi_I \sqrt{ L^4 \sin^4\psi_I \sin^2 \theta_I +(B^{NS}_{\theta \varphi})^2  }\ .\label{D5DBI}
\end{align}
Let us consider the form of $B^{NS}_{\theta \varphi}$. We consider a background field such that the following integral is constant:
\begin{equation}
r=\int _{ \mathbb{S}^2} B_2^{NS}\label{2formB}\ .
\end{equation}
We can clearly think in terms of the analogy of a magnetic monopole being placed at the origin. Therefore, we can assume the following form, 
\begin{eqnarray}
 B_2^{NS}={r\over 4\pi} \sin \theta_I d \theta_I \wedge d \varphi_I \  \longrightarrow \ \int_{ \mathbb{S}^2} {r\over 4\pi} \sin \theta_I d \theta_I \wedge d \varphi_I={r\over 4\pi} \int d\theta_I d\varphi_I \sin \theta_I=r\ .
\end{eqnarray}
It is consistent with our assumption \eqref{2formB}. Futhermore, if we define
\begin{eqnarray}
 B^{NS}_{\theta \varphi}={r\over 4\pi} \sin \theta_I\equiv b_{NS} \sin \theta_I\ ,
\end{eqnarray}
we can write the integral \eqref{D5DBI} as 
\begin{eqnarray}
S_{D5} =-T_{D5} \int d^4 x \int d\theta_I d \varphi_I\, \sin \theta_I \sqrt{L^4 \sin^4\psi_I  +\Big({r\over 4\pi}\Big)^2}=-T_{D5} A(\psi_I) \int d^4 x\ , \label{DecayPure}
\end{eqnarray}
where $A(\psi_I)=4\pi  \sqrt{L^4 \sin^4\psi_I +\left(\frac{r}{4\pi}\right)^2}$. It is evident that shifting $\Psi_I$ slightly from zero or $\pi$ increases energy. This tendency implies there exists a potential barrier between D5-branes and anti D5-branes.

We can see that if $B^{NS}_{\theta \varphi}=0$, taking $L\to 0$, the energy is zero, whereas if $B^{NS}_{\theta \varphi}\neq 0$, the energy is non-zero even if $L=0$. If there is no kinetic term, then $L=-V$, so the energy is this action multiplied by a minus. Writing The potential density at $L=0$ as $V_\star$, we obtain
\begin{align}
V_\star =rT_{D5} \ , \quad r\equiv  \int _{ \mathbb{S}^2} B_2^{NS}\ .
\end{align}
On the one hand, when we wrap $N$ branes on top of each other, $V_\star$ is simply $N$ times larger. On the other hand, what about the case where we wrapped antibranes around? When there are only antibranes, $V_\star =r T_{D5}$, since it is the same as the case when there are normal branes. Now let us consider a situation where $N+n$ brains and $N$ antibranes are wrapped around. In this case, the vacuum energy is expressed as
\begin{align}
V_{\rm meta}=(N+n) V_\star +n V_\star\ .
\end{align}
In contrast, the lowest energy state is the state where $N$ pairs of branes disappear, and $N$ branes remain, 
\begin{eqnarray}
V_{\rm true}=N V_\star \ ,
\end{eqnarray}
which we want to take as a reference. Therefore, we can add $-NV_\star$ to the constant part of the potential. The energy density of the metastable state is then
\begin{eqnarray}
\widetilde{V}_{\rm meta}= (N+n) V_\star +n V_\star- NV_\star=2n V_\star\ .
\end{eqnarray}
For simplicity, we will assume $N=0$ and $n=1$.

Now, let us look at the behavior of the system when a bubble is being created, and the vacuum is decaying. In the absence of a catalytic D3-brane, the energy of a D5-brane wrapped around $ \mathbb{S}^2$ with $\psi_I=0$ can be simply expressed by using the previous result, \eqref{DecayPure}, as
\begin{eqnarray}
E_{D5}=T_{D5} r \int d^3 x\ .
\end{eqnarray}
Calculating the energy of a bubble of radius $R$ in the Minkowski spacetime, we should get
\begin{eqnarray}
\Delta E_{D5}=T_{D5} r \Big(\int d^3x -{4\over 3}\pi R^3\Big) \ . \label{Epure}
\end{eqnarray}
Consider wrapping a D3-brane around the internal space (see Figure 13). The D3-brane, which is attached to the D5-brane, dissolves and can be regarded as a magnetic monopole on the D5-brane. Due to this effect, we must reconsider the DBI action with the magnetic field in Minkowski spacetime. The discussion about the internal space is sufficient, but the Minkowski part of the induced metric needs to be changed as follows
\begin{eqnarray}
\eta_{\mu \nu} \longrightarrow \begin{pmatrix} 
 -1 &  0 & 0 &0 \\ 
 0&1& 0   & 0 \\ 
 0&0& \xi^2   & b_{D3}\sin \theta \\ 
  0&0& - b_{D3} \sin \theta & \xi^2 \sin^2 \theta \\ 
   \end{pmatrix}
   \ ,
\end{eqnarray}
where we denote the spacial coordinates as $(\xi, \theta ,\varphi)$.
Note that $2\pi \alpha^\prime F_{\theta \varphi}=b_{D3}\sin \theta$ is the magnetic field on the brane and is different from the 2-form gauge field $B_2^{NS}$ mentioned above. Then, the energy without the hole is altered to 
\begin{eqnarray}
E^{\rm before}_{D5}=T_{D5} r \int_0^\infty d \xi \Big( 4\pi \sqrt{\xi^4+b_{D3}^2}\Big) \ .
\end{eqnarray}
Calculating the energy with a hole of radius $R$, we now have
\begin{eqnarray}
 E^{\rm after}_{D5}=T_{D5} r \int_0^\infty d \xi \Big( 4\pi \sqrt{\xi^4+b_{D3}^2}\Big) - T_{D5} r \int_0^R d \xi \Big( 4\pi \sqrt{\xi^4+b_{D3}^2}\Big) \ ,
\end{eqnarray}
and $R$-dependence is different from \eqref{Epure} when the bubble radius is small. In \cite{Kasai:2015exa}, the authors discussed this in the limit of approximately large bubbles, but here we use the above expression to refine the description of smaller bubbles. Also, if $R\to \infty$, then $E^{\rm after}_{D5}$ is zero, which means that the energy of the disappeared vacuum is now zero. Therefore, taking the difference between $E^{\rm before}_{D5}$ and $E^{\rm after}_{D5}$, we can write the obtained energy $\Delta E>0$ as
\begin{eqnarray}
\Delta E= T_{D5} r \int_0^R d \xi \Big( 4\pi \sqrt{\xi^4+b_{D3}^2}\Big)=T_{D5} r  b_{D3} \Big[4\pi R\cdot {}_2F_{1} \Big(-{1\over 2}, {1\over 4}, {5\over 4},-{R^4\over b_{D3}^2} \Big) \Big]\ ,
\end{eqnarray}
where $ {}_2F_{1} $ is the hypergeometric function.

\begin{figure}[H]
\centering
\hfill
\begin{minipage}[htbp]{0.45\linewidth}
    \begin{center}
    \includegraphics[width=0.9\linewidth]{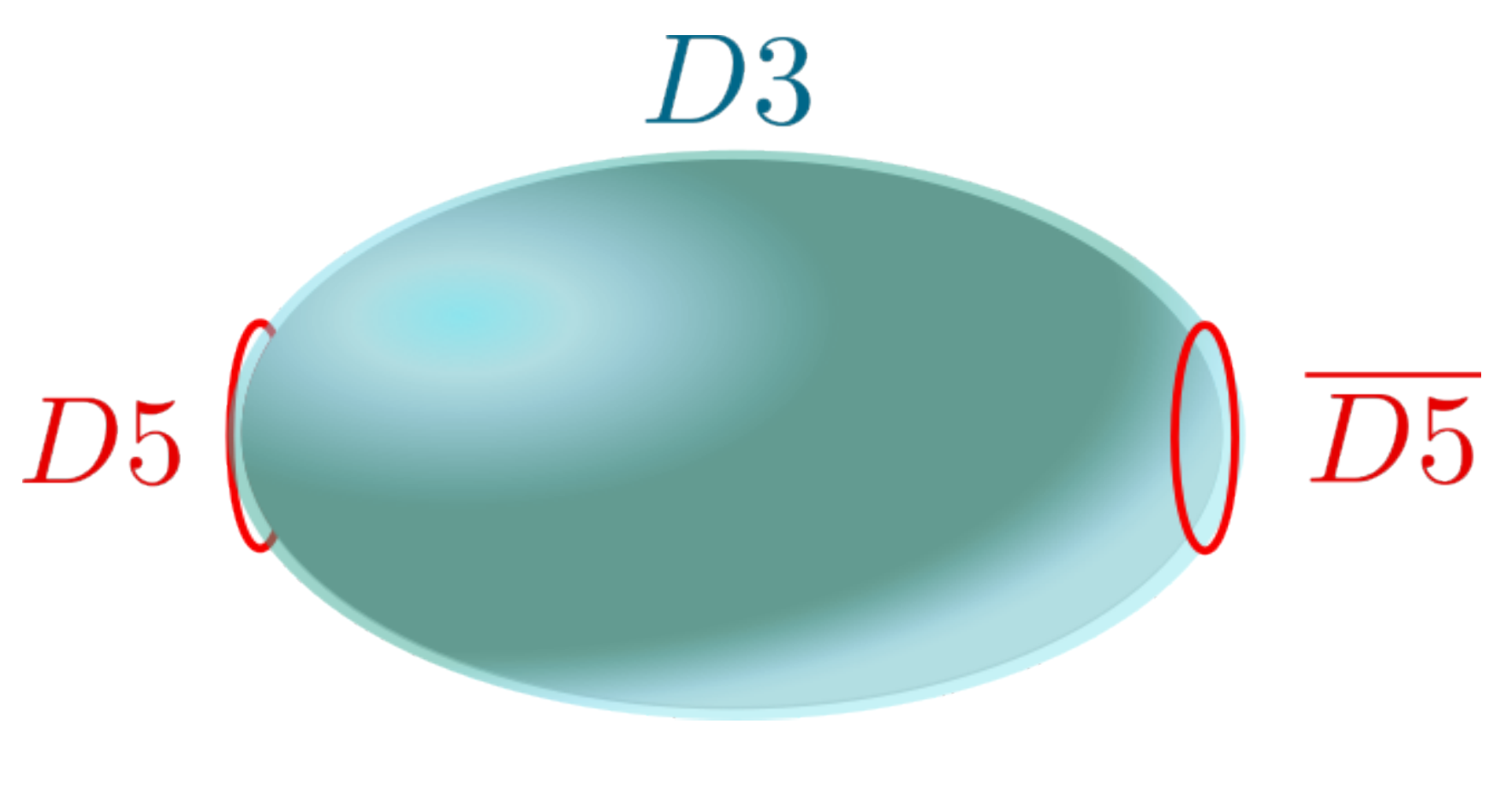}
    \vspace{-.0cm}
\end{center}
\end{minipage}
\begin{minipage}[htbp]{0.45\linewidth}
    \begin{center}
    \includegraphics[width=0.65\linewidth]{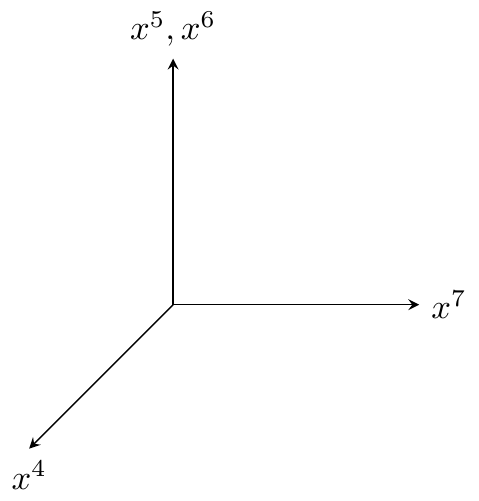}
    \vspace{-.0cm}
\end{center}
\end{minipage}
\caption{\sl The $D3$ brane as a catalyst is wrapped around $ \mathbb{S}^3$.
    }
\end{figure}

Next, we will consider the contribution from the domain wall that connects the brane and the antibrane. Since the domain wall is the D5-brane wrapped around the internal space, the three spatial dimensional is wrapped around the internal space. Since this part is unaffected by bubble size changes, we can consider $T_{DW}=T_{D5}V_3$. The remaining two spatial dimensions are the sphere of the bubble in Minkowski spacetime. Referring to the example above, the embedding function for this part is
\begin{align}
X^0 =t \ , \quad X^1 =R(t) \sin \theta \cos \varphi \ , \quad X^2= R(t) \sin \theta \sin \varphi \ , \quad X^3=R(t)\cos \theta\ . \nonu
\end{align}
This part of the calculation has the same form as the matrix in which $ R$ replaces $\xi$. The contribution of the internal space is also the same as before, so we end up with
\begin{align}
S_{\rm DW}= -T_{D5}4\pi \int d t   \sqrt{(R^4+b_{D3}^2)(1-\dot{R}^2)} \cdot \Big[2\pi^2 L^3 \int_0^\pi d \psi_I \, {2\over \pi}\sqrt{\sin^4 \psi_I +\Big({b_{NS}\over L^2}\Big)^2 } \Big]\ .
\end{align}
If we define the tension of the domain wall, $T_{\rm DW}$, as 
\begin{align}
T_{DW}=T_{D5}\Big[2\pi^2 L^3 \int_0^\pi d \psi_I \, {2\over \pi}\sqrt{\sin^4 \psi_I +\Big({b_{NS}\over L^2}\Big)^2 } \Big]\ ,
\end{align}
we can rewrite the above action as
\begin{align}
S_{DW}= -T_{DW} 4\pi \int d t   \sqrt{(R^4+b_{D3}^2)(1-\dot{R}^2)}\ ,
\end{align}
where the Lagrangian $L_{DW}$ is given by 
\begin{align}
L_{ DW}= -T_{DW} 4\pi   \sqrt{(R^4+b_{D3}^2)(1-\dot{R}^2)}\ .
\end{align}

Let us combine all the arguments so far and consider the Lagrangian when the bubble radius is $R$. Since the brane and the antibrane give similar contributions, the final action we looked for is
\begin{eqnarray}
L_{\rm total}&=& L_{DW}  +2\times L_{D5}^{\rm after} \nonu \\
&=& -T_{DW} 4\pi   \sqrt{(R^4+b_{D3}^2)(1-\dot{R}^2)}+2T_{D5} r  b_{D3} \Big[4\pi R\cdot {}_2F_{1} \Big(-{1\over 2}, {1\over 4}, {5\over 4},-{R^4\over b_{D3}^2} \Big) \Big]+\cdots\ , \nonu
\end{eqnarray}
where the ellipsis denotes any terms independent of $R$.


\section{Proof for linear relation between zero mode and time derivative of the bounce solution}

In section 3.3, we have taken the discussion with the following relation
\begin{equation}
\label{Nb2}
\widetilde{R}_{0}=\frac{1}{\sqrt{N_{b}}}\frac{d\widetilde{R}_{b}}{ds},
\end{equation}
in mind. We give simple proof here. Zero mode is one of the quantum fluctuations around the bounce solution, and we are free to choose it to be an eigenfunction of the second variation for the Euclidean action. The second variation around the bounce solution is given by
\begin{align}
\delta^2 \widetilde{S}&=\int ds\left\{\frac{\partial^2 \widetilde{L}}{\partial \widetilde{R}^2}\left(\delta \widetilde{R} \right)^2+2\frac{\partial^2 \widetilde{L}}{\partial \widetilde{R} \partial \dot{\widetilde{R}}}\delta \widetilde{R} \delta \dot{\widetilde{R}}+\frac{\partial^2 \widetilde{L}}{\partial \dot{\widetilde{R}}^2}\left(\delta \dot{\widetilde{R}} \right)^2 \right\} \nonumber \\
&=\int ds \left\{\frac{\partial^2 \widetilde{L}}{\partial \widetilde{R}^2}\left(\delta \widetilde{R} \right)^2+\frac{\partial^2 \widetilde{L}}{\partial \widetilde{R} \partial \dot{\widetilde{R}}}\delta \widetilde{R} \delta \dot{\widetilde{R}}- \frac{\partial^2 \widetilde{L}}{\partial \widetilde{R} \partial \dot{\widetilde{R}}}\delta \dot{\widetilde{R}} \delta \widetilde{R}-\frac{d}{ds}\frac{\partial^2 \widetilde{L}}{\partial \widetilde{R}\partial \dot{\widetilde{R}}}\left(\delta \widetilde{R} \right)^2\right. \nonumber \\
&\left.\hspace{7cm}-\frac{d}{ds}\frac{\partial^2 \widetilde{L}}{\partial \dot{\widetilde{R}}^2}\delta \dot{\widetilde{R}} \delta \widetilde{R} -\frac{\partial^2 \widetilde{L}}{\partial \dot{\widetilde{R}}^2}\delta \ddot{\widetilde{R}} \delta \widetilde{R} \right\} \nonumber \\
&=\int ds \left\{-\frac{d}{ds}\frac{\partial^2 \widetilde{L}}{\partial \dot{\widetilde{R}}^2} \left(\frac{d}{ds}\delta \widetilde{R}\right) \delta \widetilde{R} -\frac{\partial^2 \widetilde{L}}{\partial \dot{\widetilde{R}}^2} \left(\frac{d^2}{ds^2}\delta \widetilde{R}\right) \delta \widetilde{R}\right. \nonumber \\
&\left. \hspace{7cm}\frac{\partial^2 \widetilde{L}}{\partial \widetilde{R}^2}\left(\delta \widetilde{R} \right)^2-\frac{d}{ds}\frac{\partial^2 \widetilde{L}}{\partial \widetilde{R}\partial \dot{\widetilde{R}}}\left(\delta \widetilde{R} \right)^2 \right\}\ ,\label{一般の第二変分}
\end{align}
where we took integrations by parts in the second line. As discussed in Sec. 3.4, if we define $P[\widetilde{R}]$ and $Q[\widetilde{R}]$ as follows
\begin{equation}
    P[\widetilde{R}]\equiv \frac{\partial^2 \widetilde{L}}{\partial \dot{\widetilde{R}}^{2}}\ ,\quad Q[\widetilde{R}]\equiv \frac{\partial^2 \widetilde{L}}{\partial \widetilde{R}^2}-\frac{d}{ds}\frac{\partial^2 \widetilde{L}}{\partial \widetilde{R}\partial \dot{\widetilde{R}}}\ ,
\end{equation}
we find that the zero mode satisfies the following equation
\begin{equation}
\label{zeromode}
\left[-\frac{d}{ds}\left(P[\widetilde{R}_{b}]\frac{d}{ds} \right)+Q[\widetilde{R}_{b}]\right]\widetilde{R}_{0}=0 \ .
\end{equation}
Since we are considering the Gaussian integration over periodic configurations in WKB analysis, fluctuation modes should also be periodic. Zero mode is no exception, satisfying the periodic boundary condition
\begin{equation}
    \widetilde{R}_{0}\left(-\frac{\beta}{2} \right)=\widetilde{R}_{0}\left(\frac{\beta}{2} \right)\ ,\quad \dot{\widetilde{R}}_{0}\left(-\frac{\beta}{2} \right)=\dot{\widetilde{R}}_{0}\left(\frac{\beta}{2} \right)\ . \label{R0BC}
\end{equation}

Next, we consider the time derivative of the bounce solution.
The bounce solution satisfies the following equation of motion,
\begin{equation}
\label{ELeq}
\left(\frac{d}{ds}\frac{\partial \widetilde{L}}{\partial \dot{\widetilde{R}}_{b}}-\frac{\partial \widetilde{L}}{\partial \widetilde{R}_{b}} \right) =0\ .
\end{equation}
Performing another derivative with respect to $s$, we obtain
\begin{gather}
\left\{-\frac{\partial^2 \widetilde{L}}{\partial \dot{\widetilde{R}}^2_{b}}\frac{d^2}{ds^2}-\frac{d}{ds}\frac{\partial^2 \widetilde{L}}{\partial \dot{\widetilde{R}}_{b}^2}\frac{d}{ds}+\frac{\partial^2 \widetilde{L}}{\partial \widetilde{R}_{b}^2}-\frac{d}{ds}\frac{\partial^2 \widetilde{L}}{\partial \widetilde{R}_{b} \partial \dot{\widetilde{R}}_{b}} \right\}\dot{\widetilde{R}}_{b}=0\ . \label{Rdot}
\end{gather}
Since the bounce solution is periodic, its time derivative should also be periodic. That is, $\dot{\widetilde{R}}_{b}$ satisfies the boundary condition \eqref{R0BC}. Therefore, $\dot{\widetilde{R}}_{b}$ is also zero mode and corresponds to $\widetilde{R}_{0}$ with the normalization coefficient as \eqref{Nb2}.

\section{Review of analytic continued spectral zeta function}

Functional determinants representing 1-loop quantum fluctuations suffer from UV divergence naively and must be regularized somehow. In this respect, zeta function regularization is the very effective and well-developed method, but its computation for general Sturm-Liouville operators has been a longstanding problem. Kirsten and MacKane developed a \textit{contour deformation method} to resolve this problem \cite{Kirsten:2003py,Kirsten:2004qv}. In this section, we review their method and the recent discussion on the spectral zeta function mainly based on \cite{Gesztesy:2017xdd,Fucci:2021gos}.

\subsection{Spectral zeta function and contour deformation method}
Let us consider a Strum-Liouville operator 
\begin{equation}
    \label{SL}
    T_{A,B}
    =\frac{1}{r(x)}\left[-\frac{d}{dx}\left(p(x)\frac{d}{dx}\right)+q(x) \right],\quad x\in (A,B)\ ,
\end{equation}
where we assume $r(x)>0$,\ $p(x)>0$ over the finite interval $(A,B)$.
$r(x)$ is a weight function, and it is suitable to fix $r(x)=1$ in our problem.
If we write $\lambda_{n}$ as an eigenvalue of this operator, the spectral zeta function is defined as 
\begin{equation}
    \label{spectreZeta}
    \zeta(s;T_{A,B})\equiv\sum_{n}\lambda^{-s}_{n}\ .
\end{equation}
It is known that this spectral zeta function can be represented by the contour integral of Fredholm trace, which is a meromorphic function that has simple poles at the nonzero eigenvalues $\lambda_{n}$ \cite[Lemma 2.6.]{Gesztesy:2017xdd}
\begin{equation}
    \label{Lemma2.6}
    \zeta(s,T_{A,B})
    =-\frac{1}{2\pi i}\oint_{\gamma}d\lambda \lambda^{-s}\left(\mathrm{tr}\left(\left(T_{A,B}-\lambda I_{\mathcal{H}} \right)^{-1} \right)+\lambda^{-1}m(0;T_{A,B}) \right)\ .
\end{equation}
The second term in the integrand corresponds to a subtraction of the zero mode contribution, and $m(0; T_{A, B})$ represents the multiplicity. The counterclockwise path $\gamma$ surrounds the eigenvalues located at the real axis of the complex $\lambda$-plane and is sunk to avoid the origin, see Figure 14.

Here we introduce characteristic functions defined by the boundary conditions \cite[Sect. 1.2]{Naimark:1967}, \cite[Sect. 3.2]{zettl:2005}. These can be divided into two categories: \textit{separated} or \textit{coupled}.
\begin{enumerate}
    \item \textit{Separated boundary condition} ~\\
    The boundary conditions at $x=a$ and $x=b$ are separately given as follows
        \begin{align}
    \begin{aligned}
        &g(a)\cos(\alpha)+g^{[1]}(a)\sin(\alpha)=0\ , \\
        &g(b)\cos(\beta)-g^{[1]}(b)\sin(\beta)=0\ ,
    \end{aligned}
    \end{align}
    where $g(x)$ is an eigenfunction of the Strum-Liouville operator.
    These boundary conditions are called \textit{separated}.
    If we define a boundary value $U_{\alpha,\beta,i}(f)$ as 
        \begin{align}
    \begin{aligned}
        &U_{\alpha,\beta,1}(f)=f(a)\cos(\alpha)+f^{[1]}(a)\sin(\alpha)\ , \\
        &U_{\alpha,\beta,2}(f)=f(b)\cos(\beta)-f^{[1]}(b)\sin(\beta)\ ,
    \end{aligned}
    \end{align}
    the characteristic function is given by
       \begin{equation}
        \label{F_separated}
        F_{\alpha ,\beta}(\lambda)
        =\det \begin{pmatrix}
        U_{\alpha,\beta,1}(\theta_\lambda) & U_{\alpha,\beta,1}(\phi_\lambda) \\
        U_{\alpha,\beta,2}(\theta_\lambda) & U_{\alpha,\beta,2}(\phi_\lambda) \\
        \end{pmatrix}
        \ ,
    \end{equation}
    where $\theta_\lambda$ and $\phi_\lambda$ are independent fundamental solutions of the Sturm-Liouville equation such that the following boundary conditions are satisfied
       \begin{equation}
        \label{thetaphi}
        \theta_\lambda(a)=\phi^{[1]}_\lambda(a)=1\ ,\quad \theta^{[1]}_\lambda(a)=\phi_\lambda(a)=0\ .
    \end{equation}
    $g^{[1]}$ represents a pseudo-derivative of $g$, which is defined as $g^{[1]}=p(x)g^{\prime}(x)$.
       \item \textit{Coupled boundary condition}  ~\\
    The boundary values at $x=a$ and $x=b$ are related as follows
        \begin{align}
        \begin{pmatrix}
        g(b) \\ g^{[1]}(b)  
        \end{pmatrix}
        =e^{i\varphi}R
        \begin{pmatrix}
        g(a) \\ g^{[1]}(a)  
        \end{pmatrix}
        \ ,
    \end{align}
    where $\varphi \in [0,\pi)$, and $R \in SL(2,\mathbb{R})$.
    This boundary condition is called \textit{coupled}.
    If we define a boundary value $V_{\varphi,R,i}(f)$ as 
       \begin{align}
    \label{V_expression}
        \begin{aligned}
            &V_{\varphi,R,1}(f)=f(b)-e^{i\varphi}R_{11}f(a)-e^{i\varphi}R_{12}f^{[1]}(a)\ , \\
            &V_{\varphi,R,2}(f)=f^{[1]}(b)-e^{i\varphi}R_{21}f(a)-e^{i\varphi}R_{22}f^{[1]}(a)\ ,
        \end{aligned}
    \end{align}
    the characteristic function is given by
        \begin{equation}
        \label{F_coupled}
        F_{\varphi ,R}(\lambda)
        =\det \begin{pmatrix}
        V_{\varphi,R,1}(\theta_\lambda) & V_{\varphi,R,1}(\phi_\lambda) \\
        V_{\varphi,R,2}(\theta_\lambda) & V_{\varphi,R,2}(\phi_\lambda) \\
        \end{pmatrix}
        \ .
    \end{equation}
    \end{enumerate}
By definition, the complex numbers $\lambda$, which the characteristic function takes the zeros, coincide with the eigenvalues of the original Strum-Liouville problem.

We can directly relate the above characteristic function to a Fredholm determinant and a Fredholm trace \cite[Thm.3.4.]{Gesztesy:2017xdd}.
Since it should be fine to adopt the periodic boundary condition for the vacuum decay, $\varphi=0$, $R=I$, we only focus on $F_{\varphi, R}(\lambda)$ in the following discussion.
Suppose $\lambda_{0}$ is one of the eigenvalue, $\lambda_{0}\in \rho(T_{\varphi,R})$, then
\begin{align}
        \mathrm{det}_{L^{2}_{r}((a,b))}\left(I_{L^{2}_{r}((a,b))}-(\lambda-\lambda_{0})\left(T_{\varphi,R}-\lambda_{0}I_{L^{2}_{r}((a,b))} \right)^{-1} \right)
        =\frac{F_{\varphi,R}(\lambda)}{F_{\varphi,R}(\lambda_{0})},\quad \lambda\in \mathbb{C}\ .
\end{align}
For the Fredholm trace, in particular, 
\begin{equation}
    \label{tr_coupled}
        \mathrm{tr}_{L^{2}_{r}((a,b))}\left(\left(T_{\varphi,R}-\lambda I_{L^{2}_{r}((a,b))} \right)^{-1} \right)
        =-\frac{d}{d\lambda}\ln \left(F_{\varphi,R}(\lambda) \right)\ ,\quad \lambda\in \rho(T_{\alpha,\beta})\ .
\end{equation}
Substituting this into \eqref{Lemma2.6}, then we can rewrite the spectral zeta function as
\begin{align}
\label{zeta_characteristic}
\begin{aligned}
    \zeta(s;T_{A,B})
    =\frac{1}{2\pi i}\oint_{\gamma}d\lambda \lambda^{-s}\left(\frac{d}{d\lambda}\ln (F_{A,B}(\lambda))-\lambda^{-1}m_{0} \right)\ ,
\end{aligned}
\end{align}
where we rewrite as $m_{0}=m(0;T_{A,B}) $.

Let us assume
\begin{equation}
    \label{branch}
    R_{\psi}
    =\left\{\lambda =te^{i\psi}|t\in [0,\infty ) \right\}\ ,\quad \psi\in(\pi/2,\pi)\ 
\end{equation}
is a branch cut of $\lambda^{-s}$.
Deforming the path to hold the branch cut, as shown in Figure 14, the contour integral replaces the calculation of the discontinuity across the branch cut
\begin{equation}
    \label{zeta_ContorDeformed}
    \zeta(s;T_{A,B})
    =e^{is(\pi-\psi)}\frac{\sin(\pi s)}{\pi}\int^{\infty}_{0}dt t^{-s}\frac{d}{dt}\ln \left(F_{A,B}(t e^{i\psi})t^{-m_{0}}e^{-im_{0}\psi} \right)\ .
\end{equation}
\begin{figure}[H]
\centering
\hfill
\begin{minipage}[htbp]{0.45\linewidth}
    \begin{center}
    \includegraphics[width=0.9\linewidth]{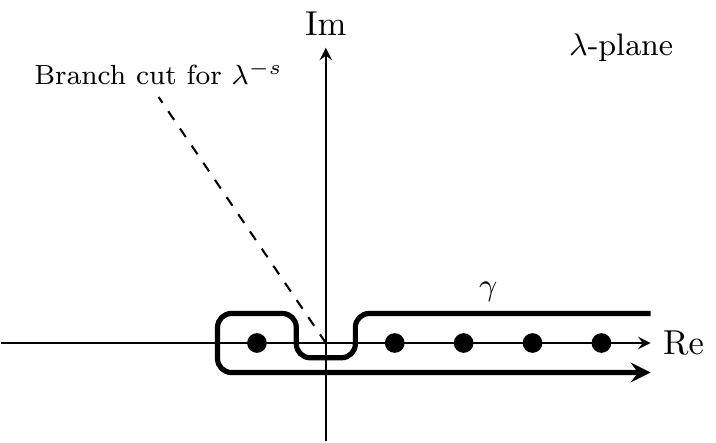}
    \vspace{-.0cm}
\end{center}
\end{minipage}
\begin{minipage}[htbp]{0.45\linewidth}
    \begin{center}
    \includegraphics[width=0.9\linewidth]{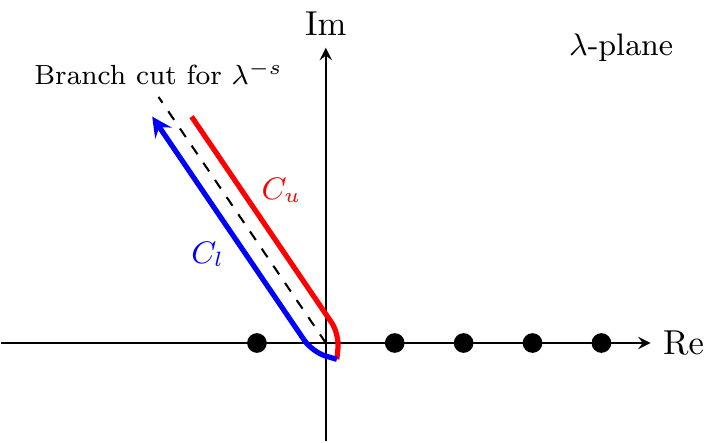}
    \vspace{-.0cm}
\end{center}
\end{minipage}
\caption{The Contours $\gamma$, $C_{u}$ and $C_{l}$ on the complex $\lambda$-plane.The dashed lines represent the branch cut of $\lambda^{-s}$. The brobs on the real axis correspond to the eigenvalues of the differential operator $T_{A,B}$.}
\end{figure}

\subsection{Derivative at $s=0$}
The calculation for the zeta regularized determinants requires the derivative at $s=0$ of the spectral zeta function, so we must take care of the regularity there.
The regular region of the spectral zeta function represented by the contour integral is determined by the asymptotic behavior of $\ln(F_{A, B}(\lambda))$ in $\lambda\rightarrow0$ and $\lambda\rightarrow \infty$.
 One the one hand, the characteristic function $F_{A,B}(\lambda)$ has a small $\lambda$ expansion \eqref{F_expansion},
\begin{equation}
    F_{A,B}(\lambda)
    =F_{m_{0}}\lambda^{m_{0}}+\sum_{n=m_{0}+1}^{\infty}F_{n}\lambda^{n}\ ,
\end{equation}
where $m_{0}$ denote the multiplicity of zero eigenvalues.
Then in the limit $\lambda \rightarrow 0$, 
\begin{equation}
    \frac{d}{d\lambda}\ln(F_{A,B}(\lambda)\lambda^{-m_{0}})
    \underset{\lambda \rightarrow 0}{=}
    O(1)\ .
\end{equation}
On the other hand, $F_{A,B}(\lambda)$ has a large $\lambda$ asymptotic expansion as 
\begin{align}
    \label{assym_lnF}
    \ln(F_{A,B}(\lambda))
    \underset{|\lambda|\rightarrow \infty,\ \mathrm{Im}(\lambda^{1/2})\geq 0}{=}
    -ic\lambda^{1/2}-\frac{k_{0}+1}{2}\ln(\lambda)+\ln\left(\frac{\Gamma_{k_{0}}}{2ic} \right)+\sum_{m=1}^{N}\Psi_{m}\lambda^{-m/2}\ ,
\end{align}
see \cite[Sect.3.2]{Fucci:2021gos} for a detailed discussion of the derivation. 
Thus, in the limit $|\lambda| \rightarrow \infty$,
\begin{align}
    \frac{d}{d\lambda}\ln(F_{A,B}(\lambda)\lambda^{-m_{0}})
    \underset{|\lambda|\rightarrow \infty,\ \mathrm{Im}(\lambda^{1/2})\geq 0}{=}
    O\left(\frac{1}{\sqrt{\lambda}} \right)\ .
\end{align}
Considering the discussion above, we find that the contour deformation method described above is well-defined only in the region of $1/2<s<1$ and not valid for $s=0$.

To perform a derivative at $s=0$, we have to perform an analytic continuation to the left of the abscissa of the boundary $s=1/2$. All one has to do is to subtract $N$ terms from the asymptotic expansion \eqref{assym_lnF} and then add them back. It means
\begin{equation}
    \label{zeta_AC}
    \zeta(s;T_{A,B})
    =Z(s,A,B)+\sum_{j=-1}^{N}h_{j}(s,A,B)\ ,
\end{equation}
The explicit forms of each function are given by
\begin{align}
    \label{zeta_Z}
    \begin{aligned}
       Z(s,A,B)=& e^{is(\pi-\psi)}\frac{\sin(\pi s)}{\pi}\int^{\infty}_{0}dt t^{-s}\frac{d}{dt}\left\{\ln \left(F_{A,B}(t e^{i\psi})t^{-m_{0}}e^{-im_{0}\psi} \right)\right. \\
       &\left.-H(t-1)\left[-ic\lambda^{1/2}e^{i\psi/2}-\left[\frac{k_{0}+1}{2}+m_{0}\right]\ln(t) \right.\right. \\
       &\left. \left.-\left[\frac{k_{0}+1}{2}+m_{0} \right]i\psi+\ln\left(\frac{\Gamma_{k_{0}}}{2ic} \right)+\sum_{n=1}^{N}\Psi_{n}e^{in\psi/2}t^{-n/2} \right] \right\}\ ,
    \end{aligned}
\end{align}
\begin{align}
    \label{zeta_h}
    \begin{aligned}
        &h_{-1}(s,A,B)=-ie^{i\pi(\pi-\psi)}\frac{\sin(\pi s)}{\pi}\frac{ce^{i\psi/2}}{2s-1}\ , \\
        &h_{0}(s,A,B)=-(k_{0}+1+2m_{0})e^{is(\pi-\psi)}\frac{\sin(\pi s)}{2\pi s}\ , \\
        &h_{n}(s,A,B)=-e^{is(\pi-\psi)}\frac{\sin(\pi s)}{\pi}\frac{n}{2s+n}e^{-in\psi/2}\Psi_{n}\ .
    \end{aligned}
\end{align}
where $H(s)$ is a step function.
In this way, the regular region of the spectral zeta function is extended to $-(N+1)/2<\mathrm{Re}(s)<1$. Since $N=0$ is sufficient to include $s=0$, using \eqref{zeta_AC}, the derivative of the spectral zeta function at $s=0$ obeys
\begin{align}
    \zeta^{\prime}(0;L_{A,B})&=Z^{\prime}(0,A,B)+h^{\prime}_{-1}(0,A,B)+h_{0}^{\prime}(0,A,B) \nonumber \\
    \label{zetaprime}
    &=i\pi n-\ln \left(2c\left|\frac{F_{m_{0}}}{\Gamma_{k_{0}}} \right| \right)\ .
\end{align}


\end{document}